\documentclass[natbib]{svjour3}                     
\smartqed  
\usepackage{graphicx}
\usepackage{bbm}
\usepackage{bbold}
\usepackage{amsmath}
\usepackage{capt-of}
\usepackage{ulem} 
\newcommand{\mbf}{\boldsymbol} 
\newcommand{\I}{\mathrm{i}} 

\newcommand{\secs}{\rm{\,seconds}}

\newcommand{\aap}{{Astron. Astrophys.}}
\newcommand{\apj}{{Astrophys. J.}}
\newcommand{\apjl}{{Astrophys. J. Lett.}}
\newcommand{\grl}{{Geophys. Res. Lett.}}
\newcommand{\solphys}{{Solar Phys.}}
\newcommand{\ssr}{{Space Sci. Rev.}}
\newcommand{\mnras}{{Mon. Not. Roy. Astron. Soc.}}
\begin{document}

\title{Recent developments in helioseismic analysis methods and solar data assimilation
}

\titlerunning{Recent developments in helioseismic analysis methods and solar data assimilation}        

\author{A. Schad\and L. Jouve \and T. L. Duvall Jr. \and M. Roth \and S. Vorontsov}


\institute{T. L. Duvall, Jr.
 \at Max-Planck-Institut f\"ur Sonnensystemforschung, Justus-von-Liebig-Weg 3, 37077 G\"ottingen, Germany 
 \and 
L. Jouve 
  \at Universit\'e de Toulouse, UPS-OMP, Institut de Recherche en Astrophysique et Plan\'etologie, 31028 Toulouse Cedex 4, France
\and 
M. Roth \and A. Schad 
  \at Kiepenheuer-Institut f\"ur Sonnenphysik, Sch\"oneckstr. 6, 79104 Freiburg, Germany
\and 
S. V. Vorontsov 
  \at Astronomy Unit, School of Physics and Astronomy, Queen Mary University of London, Mile End Road, London E1 4NS, UK; and
Institute of Physics of the Earth, B. Gruzinskaya 10, Moscow 123810, Russia
}

\date{Received: date / Accepted: date}

\maketitle

\begin{abstract}
We review recent advances and results in enhancing and developing helioseismic analysis methods and in solar data assimilation. In the first part of this paper we will focus on selected developments in time-distance and global helioseismology. In the second part, we review the application of data assimilation methods on solar data. Relating solar surface observations as well as helioseismic proxies with solar dynamo models by means of the techniques from data assimilation is a promising new approach to explore and to predict the magnetic activity cycle of the Sun.

\end{abstract}

\section{Introduction}
\label{intro}
Helioseismic inferences are subject to statistical and systematic errors of different origin. Part of the errors may be reduced by analyzing longer observations or averaging over observations. But especially systematic errors often result from the neglect of inevitable instrumental effects on the observations or from model misspecifications in the data analysis and cannot be diminished for example by longer observations. Consequently, more accurate models and a better account of instrumental effects are essential for the enhancement of local and global helioseismic analysis methods. This will improve the accuracy and reliability of estimates of helioseismic parameters like mode frequencies and wave travel times. Just recently, for example, progress was made in measuring the meridional flow in the deep layers with the time-distance technique not only by the availability of high-resolution data from the HMI instrument but also after becoming aware and removal of a systematic center-to-limb effect in the travel--time estimates~\citep{Zhao12,Zhao13}. In order to detect weak processes like the meridional flow in the deeper interior or highly dynamic processes like supergranular convective motions with the time-distance technique, travel--time measurements of high signal-to-noise ratio are needed. The development of spatial averaging strategies and optimized filters can help here. Regarding global helioseismic investigations, different models and estimation schemes are used to extract the mode parameters of modes of medium and large harmonic degree from the data of GONG, MDI or HMI~\citep[e.g.,][]{Anderson90,Schou92,Schou98,Hill96,Korzennik05,Reiter15}. Recent advances are made on identification and eradication of systematic influences of different origin on the mode parameter estimates~\citep{Larson08,Rabello08,Vorontsov09,Korzennik13}. Further advances in global helioseismology are made by the development of analysis methods that exploit the perturbation of mode eigenfunctions due to the advection of the acoustic waves by flows. This kind of perturbation is also described as mode coupling. It manifests in correlations that can be investigated by cross-spectral analysis of spherical harmonic decomposed global oscillation data~\citep{Woodard00,Schad13}. Such an approach is promising for inferences on the meridional flow inside the Sun since its influence on mode frequencies is only of second order and very small~\citep{Roth08,Gough10,Schad11,Vorontsov11}. 

One essential purpose of helioseismic investigations of dynamic processes within the solar interior is to explore and better understand the solar dynamo and the related activity cycle. In particular, some dynamo models predict a strong link between the meridional flow speed and the magnetic cycle period. Enabling observations as well as helioseismic constraints and models to work hand in hand could definitely help to progress on our understanding of the solar cycle and possibly produce reliable forecasts of solar phenomena and activity. The concepts of data assimilation are promising for solving these issues. Data assimilation has been extensively used for decades to predict the weather on Earth. Moreover, by combining physically-based models and well-chosen observables, it allows to constrain processes unaccessible to any measurement through their indirect impact on observations. With the increasing collection of solar data, this technique has become more and more attractive in solar physics and starts to be adapted to this field.

In the following we review some of the limitations related with the estimation of the helioseismic quantities in time-distance and global helioseismology and ideas and current attempts to accomplish them. For the time-distance analysis two recent advances, ensemble averaging and optimal filtering, will be discussed in detail. The inclusion of enhanced models of the solar oscillation spectrum to improve global helioseismic parameter estimates is addressed by the ''Global Helioseismic Metrology'' project. As another advancement in global helioseismology we review the mode eigenfunction perturbation analyses that uses cross-spectra of solar oscillations. In the second part of this paper we overview the latest developments and results from application of data assimilation methods in solar physics with regard to activity cycle predictions. 

\section{Developments in local helioseismology}
One recent challenge in local helioseismology is to obtain precise three-dimensional maps of the temporal evolution of the solar sub-surface layers. Here we focus on the time-distance method which is one tool to obtain such measurements and promises for example the detection of the emergence of magnetic flux or the analysis of convective motions like supergranulation.

\subsubsection{Large and small separations}
\begin{figure*}
\begin{center}
 \includegraphics[width=.9\textwidth]{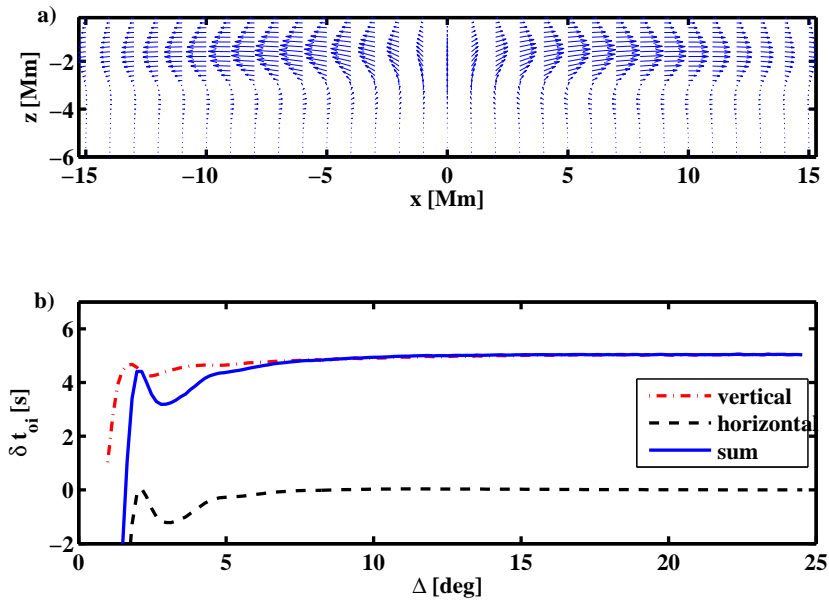}
\end{center}
\caption{a) Flow vectors of a shallow supergranular model~\citep{Duvall12}, and b) Vertical flow and horizontal flow contributions to the center-annulus
travel--time difference $\delta\rm{t_{oi}}[s]$ and their sum.}
\label{fig:duvall2}       
\end{figure*}

In time-distance helioseismology, travel times are measured between pairs of surface locations with angular heliocentric separation $\Delta$. A problem in the analysis of travel times is that information from larger depths is only obtained from signals measured at locations with larger separations $\Delta$ and these are adversely affected by larger noise levels due to the geometrical spreading of the wavefronts~(\citealt{Gizon04}, eq. (31)). In recent years much of time-distance helioseismology has been carried out using what we will call small separations, like $\Delta<2.4^\circ(29.2\,\mathrm{Mm})$~\citep{Jackiewicz08} and $\Delta<5^\circ(61\,\mathrm{Mm})$~(\citealt{Zhao03}). In particular for the analysis time periods appropriate for studying near-surface features, say 8 hours, and to get sufficient signal-to-noise ratio to easily see perturbation signals, analyses have been restricted to relatively small separations $\Delta$. One of the difficulties with the restriction to small separations $\Delta$ is the inability to separate horizontal and vertical flow signals near the surface. The travel--time signal from a flow is measured from the time difference between counterpropagating waves traveling between two points. The ray theory predicts that the time difference $\delta\tau$ is given by the integral of the flow along the ray path $\Gamma$, $\delta\tau=-2\int_{\Gamma}{\bf v}\cdot{\bf{ds}}/c^2$, where ${\bf v}$ is the flow velocity, ${\bf ds}$ is the element of length along the ray path, and $c$ is the sound speed. When studying near-surface phenomena, like supergranulation or sunspots, the restriction to short distances means that horizontal and vertical flows are not cleanly separated~\citep{Zhao03}. However, for large distances, the separation is much cleaner. To illustrate this point, Fig.~\ref{fig:duvall2}b shows the travel--time contributions of the vertical and horizontal flow components as well as the sum for the shallow supergranular flow model shown in Fig.~\ref{fig:duvall2}a. For such a shallow flow model, the horizontal component contributes almost nothing for $\Delta>6\deg$. But the vertical contribution of $\approx5[s]$ at large $\Delta$ can be used to define a constant of the model, namely an integral over depth of the vertical flow at cell center, $\delta\tau^z=-2\int{v_z \rm{dz}/c^2}$. 

To make full use of large travel distances, the signal-to-noise ratio in the travel--time maps needs to be increased. This requires several steps: optimal choice of phase-speed filters and spatial averaging schemes. For example, \cite{Ilonidis13} used a broad phase speed filter with a non-Guassian shape and particular averaging schemes with multiple arc configurations to obtain their results on the detection of emerging magnetic flux \citep{Ilonidis11}. However, the reliability of their findings still needs to be validated \citep{Braun12}. 

\subsubsection{Averaging over features}
Different averaging schemes are applied in time-distance helioseismology to get a significant signature of a particular feature in the wave travel times. As an example~\citet{Birch10} employed numerical models to estimate that a rising flux tube causes travel--time shifts in the order of 1\,s. A signal which could be certainly detected only by averaging over 150 of such events.

\begin{figure*}
\begin{center}
  \includegraphics[width=0.75\textwidth]{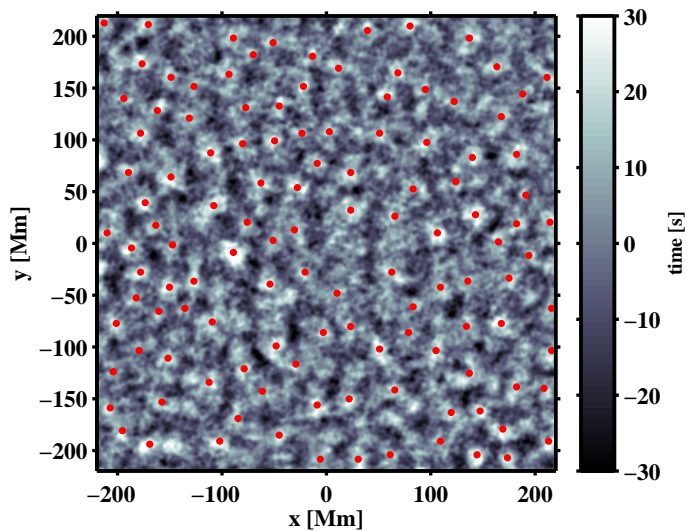}
\end{center}
\caption{Feature locations (red) determined by the local peaks in the horizontal divergence signal (gray scale) derived as the annulus minus center travel--time difference for the f mode (Figure adapted from~\citealt{Duvall10}).}
\label{fig:duvall1}       
\end{figure*}

A powerful method to increase the signal-to-noise ratio and to obtain average properties directly is to average either cross correlations or travel times about the locations of features detected near the surface. The basic assumptions of this feature averaging are that there is an underlying linear process that can be averaged and that the systematic errors are sufficiently understood. This averaging technique has been used with small magnetic features (\cite{Duvall06} and \cite{Felipe12}) and to supergranulation (\cite{Birch06}, \cite{Hirzberger08}, \cite{Duvall10}, \cite{Duvall12}), and \cite{Svanda12}). An example of features corresponding to the supergranulation cell centers is shown in Fig.~\ref{fig:duvall1} from \cite{Duvall10}. To derive these feature locations, a map of the center-annulus travel--time differences of f-mode waves is used to approximate the horizontal divergence of the flow. Local maxima of a smoothed version of this map are located and only features farther apart than 22 Mm are accepted. The smaller (in terms of travel--time difference) of a pair closer than 22 Mm is rejected. This results in a useful choice of supergranules, although it is likely biased towards larger than average cells (\cite{Svanda12}).  

From the feature analysis of \cite{Duvall10}, only a single parameter
is derived from each cell, namely the strength of the f-mode divergence
signal. Travel times could be binned based on the value of this
parameter, enabling some discrimination of different strength or size
cells. No one has published such results yet, but it is an obvious
extension of the present work. However, it is likely that an
additional parameter (at least one) would be required to describe
supergranular cells. Cell size comes to mind as a candidate. However,
it is very likely that the peak f-mode divergence signal will be 
highly correlated with the size. If the cells were to be described
by two parameters, it would be useful to have parameters that are
orthogonal. A way to analyze the horizontal divergence map to derive additional
cell information is the Fourier segmentation procedure developed by 
\cite{Hirzberger08}.  

\subsubsection{Optimized filters}
An essential element of time-distance helioseismic methods is the construction and usage of optimized filters. They have been used in time-distance helioseismology since its  inception (\cite{Duvall93}), where it was shown that filtering the data in horizontal phase speed $\omega/k$ ($\omega$ circular frequency of a wave and $k$ its horizontal wavenumber) leads to isolated features in a time-distance correlation function. This is important because waves with the same horizontal phase speed travel to the same depth in the Sun (\cite{Duvall82}).
In the early work, it was considered that by measuring travel--time 
differences between surface points that the effect of perturbations
along the ray path could be measured just by considering waves with 
the particular phase speed corresponding to that depth.  However,
it was shown by \cite{Woodard98} that perturbations, such as 
supergranulation, spread power in the $k-\omega$ power spectrum by
an amount corresponding to their inverse size.  Supergranulation,
with a spectrum peaking at spherical harmonic degree 
$\ell=kR=120$, spreads signal considerably in the power spectrum.

\begin{figure*}
\begin{center}
  \includegraphics[width=0.9\textwidth]{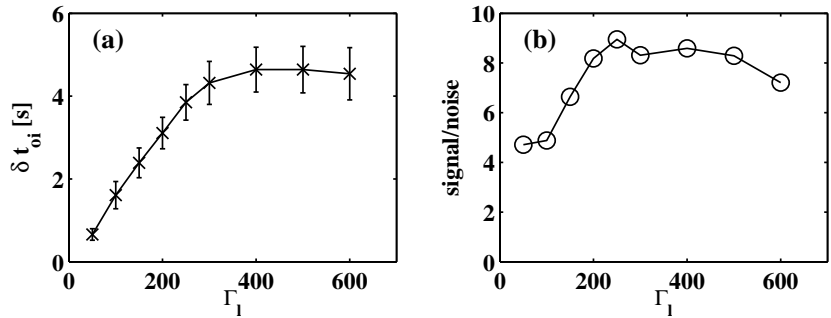}
\end{center}
\caption{(a) Travel--time difference $[\delta t_{\rm{oi}}]$ versus filter FWHM $[\Gamma_{\ell}]$. The unfiltered case has $5.3\pm1.2\secs$. (b) The travel--time difference from (a) divided by the size of the error bar from (a) {\it versus} the filter FWHM $\Gamma_{\ell}$. The value for the unfiltered case is 4.6. Figure from~\cite{Duvall12}.} 
\label{fig:duvall3}
\end{figure*}

To study features such as supergranulation, it is useful to have a
filter broad enough to admit most or all of the signal. This issue
has been examined by \cite{Duvall12}. They have constructed filters with a central phase speed but a full width
at half maximum (FWHM) $\Gamma_{\ell}$ that is independent of $\omega$. For a range of large separations of $\Delta=19-22\deg$, they have 
measured the center-annulus travel--time difference $[\delta t_{\rm{oi}}]$
for the average supergranules versus filter width
(Fig.~\ref{fig:duvall3}a). At small widths, the signal strength is
approximately linear with filter width until most of the supergranular
signal is captured for $\Gamma_{\ell}>240$. Another important parameter of 
the filter is the subsequent signal to noise ratio. This ratio is
shown in Fig.~\ref{fig:duvall3}b. It is found that narrow filters,
in addition to not capturing most of the signal, do not yield the
best signal to noise ratio. From Fig.~\ref{fig:duvall3}b a filter
width of $\Gamma_{\ell}=400$ was chosen for further study.

\subsection{Discussion}
In this section we have touched on two relatively new advances in time-distance helioseismology: ensemble or feature averaging and optimal filtering. Both strategies improve the signal-to-noise ratio of travel--time measurements that is necessary to measure the solar subsurface velocities. Of course other methods such as the ring-diagram analysis \citep{hill88}, helioseismic holography \citep{lindsey90}, and Fourier-Hankel analysis \citep{Braun87} can provide independent views on the processes insides the Sun. However, all local helioseismic techniques require a certain strategy of averaging or filtering to increase the signal-to-noise ratio. Another issue of time-distance analysis concerns its reliability and ability to retrieve subsurface structures, especially in the presence of magnetic fields~\citep{Gizon09,Moradi10}, but also of subsurface supergranular-like flows as shown by~\cite{DeGrave14}. The authors of that simulation study showed that current time-distance techniques were not able to adequately retrieve the supergranular flow pattern and suggested that averaging schemes, as proposed by~\cite{Duvall10} and \cite{Svanda11}, might help to obtain reliable results in this case.

\section{Developments in global helioseismology}
Global helioseismic analyses of mode frequencies were very fruitful in the past for our current picture of the Sun's internal structure and dynamics. The fidelity and resolution of these findings is restricted by the quality of the observations but also by systematic errors entering the analysis methods used to determine the mode parameters from the data. Especially the systematic errors are crucial for global helioseismic inferences and cannot be overcome by improving the quality or length of observations. In the following we focus on some of the difficulties in estimating global mode parameters and the efforts on identifying and eradicating associated systematic errors by means of the ``Global Helioseismic Metrology'' project. 

\subsection{Measurement of solar oscillation frequencies: Uncertainties and limitations}
\label{sec:global}
\begin{figure*}
\begin{center}
\includegraphics[width=1.0\linewidth]{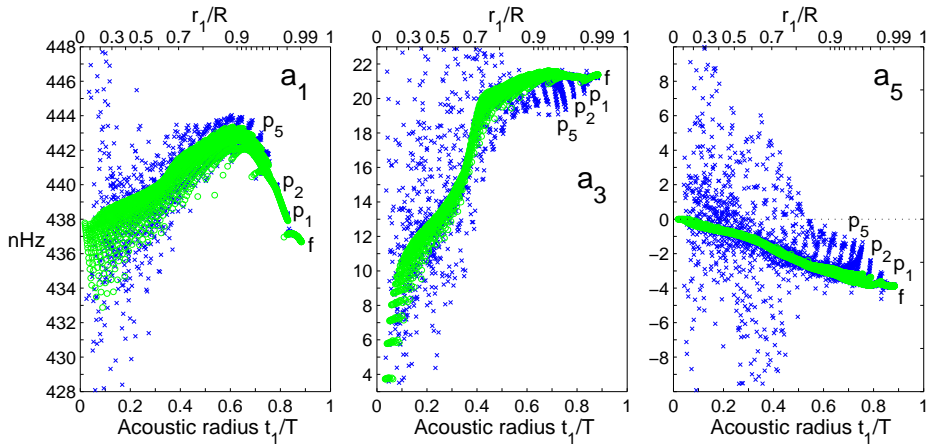}
\end{center}
\caption{First three rotational splitting coefficients $a_1, a_3, a_5$ of solar $p$- and $f$ modes, plotted versus the position of their inner turning points. Blue crosses: average of splitting coefficients measured from 27 consecutive 72d slots of SOHO MDI observations and corrected for the temporal variation of the solar rotation. Green circles: the same splitting coefficients but calculated from the inverted rotating model.}
\label{fig1svv}
\end{figure*}

\begin{figure*}
\begin{center}
\includegraphics[width=0.45\linewidth]{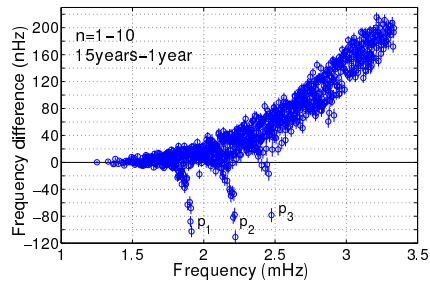}
\end{center}
\caption{Difference between centroid frequencies averaged over 15 years of SOHO MDI observations and frequencies of the 1-yr measurement at solar activity minimum~\citep{Vorontsov13a}. The error bars are 1-$\sigma$ errors of the 1-yr data set.}
\label{fig2svv}
\end{figure*}

A significant part of the uncertainty in global helioseismic measurements has been identified as systematic in nature. The evidence for these errors comes both from the direct comparison of the results provided by different data-analysis techniques applied to the same data (\cite{Larson08}; \cite{Vorontsov09}), and from the helioseismic inversions, where they are revealed as an internal inconsistency in the input data set. Fig.~\ref{fig1svv} illustrates systematic errors in published SOHO MDI rotational splitting coefficients, revealed by the inversion in an attempt to implement the splitting measurements, accumulated over five years of observations and corrected for temporal variations in the solar internal rotation (``torsional oscillations''), to improve the measurement of the time-independent component of the rotation \citep{Vorontsov02a}. The prominent horn-like structures in the mismatch between the data and the inverted model do not allow the measurement to benefit from the prolonged observation. Fig.~\ref{fig2svv} illustrates a simpler example with centroid ($m$-averaged) frequencies. Here, the centroid frequencies tabulated over the entire span of MDI observations were averaged without any corrections for variations with solar activity, for addressing the equation of state of solar plasma \citep{Vorontsov13a}. The difference between the averaged frequencies and those measured in the first year of SOHO mission (low solar activity) demonstrates a well-known frequency dependence, but only on average: the obvious outliers are the higher-degree modes, the most precious part of data for this particular measurement (modes with turning points in the vicinity of HeII ionization region). Again, attempting to implement a prolonged observation brings no benefit, but only makes systematic errors more evident. 

\begin{figure*}
\centering
\includegraphics[width=0.48\linewidth]{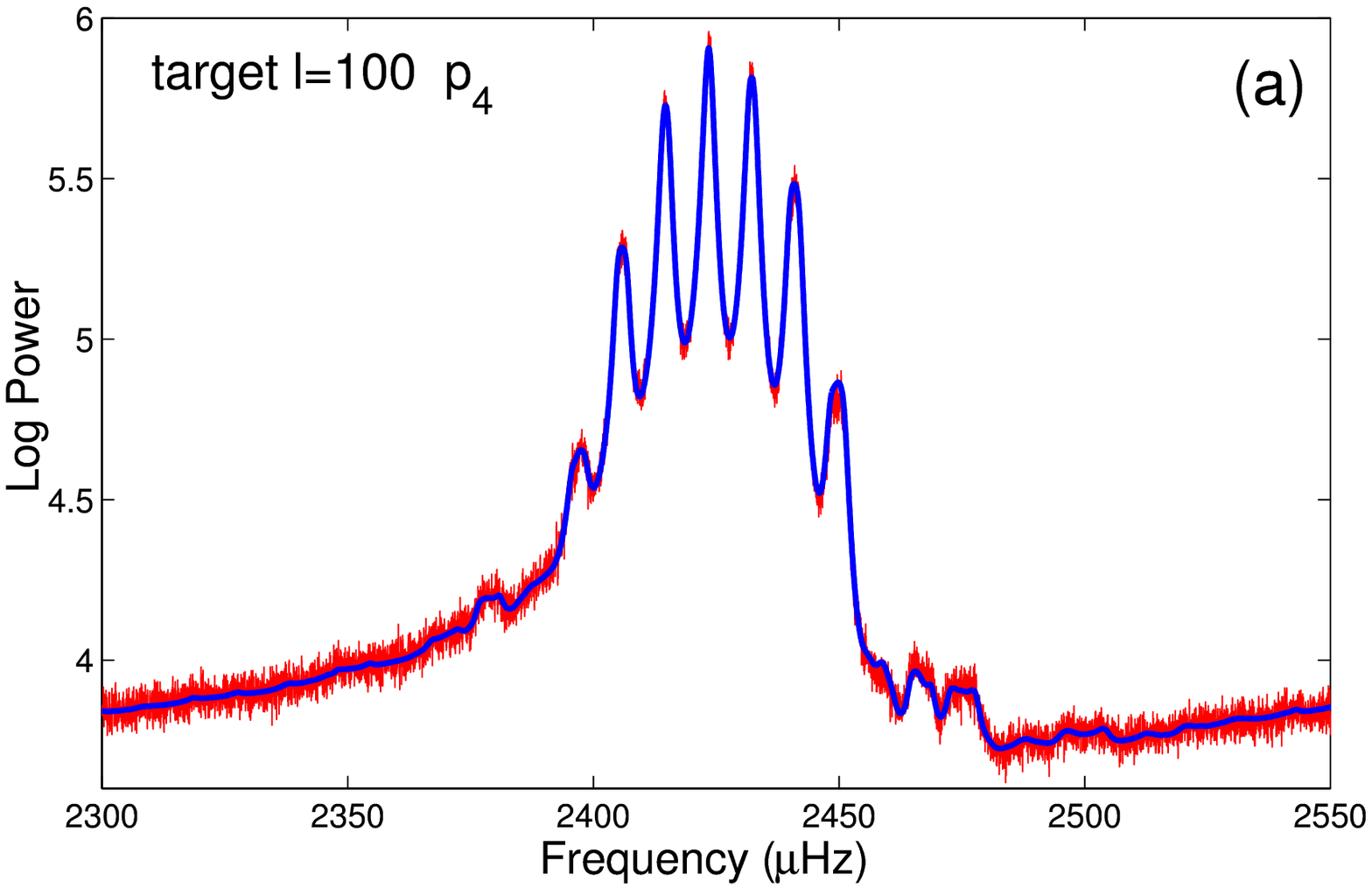}
\hspace{0.1in}%
\includegraphics[width=0.48\linewidth]{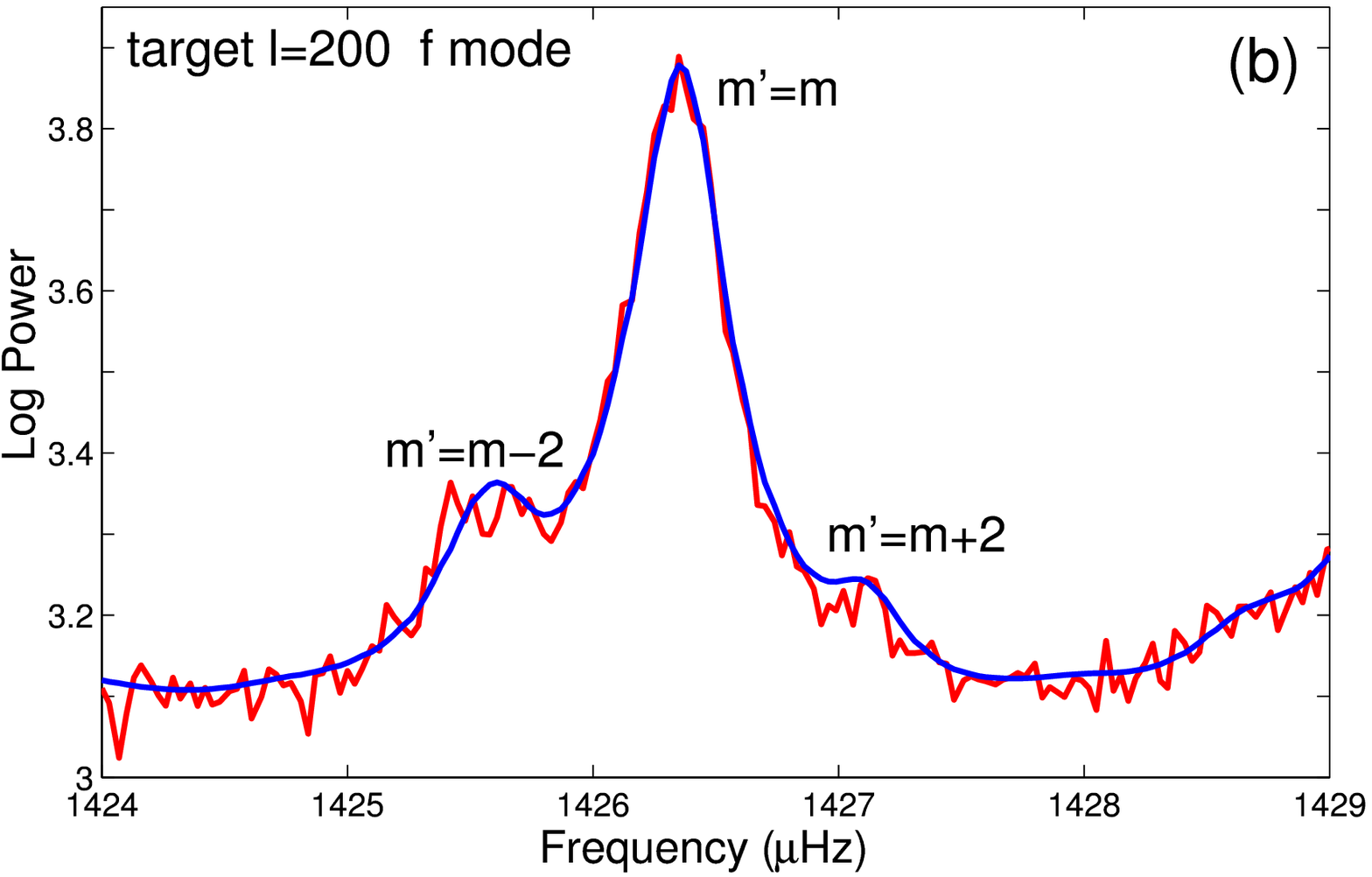}
\caption{1-year SOHO MDI power spectra (red) and their models (blue) for $p_4$ mode of degree $l=100$ (a) and $f$ mode of $l=200$ (b).}
\label{fig3svv}
\end{figure*}

The challenges to the accurate measurement of the solar oscillation frequencies are illustrated by two examples of Doppler-velocity power spectra shown in Fig.~\ref{fig3svv}. The difficulties originate largely from the spatial leaks coming from modes of neighboring values of degree $l$ (Fig.~\ref{fig3svv}a) and azimuthal order $m$ (Fig.~\ref{fig3svv}b). The accurate modeling of the spatial leaks, which may blend into the target peak, puts very tough requirements on the accuracy of the leakage matrix. Further, the asymmetry of the line profiles (Fig.~\ref{fig3svv}a) needs to be properly accounted for. Finally, deviations from the spherical symmetry of the equilibrium solar configuration lead, in general, to mode coupling---instead of individual modes described by a particular pair of $(l,m)$ values, we have the coherent composite states. The biggest effect comes from the mode coupling by differential rotation; an example of how the mode coupling affects the power spectra is the asymmetry of the amplitudes of $m-2$ and $m+2$ leaks, clearly seen in Fig.~\ref{fig3svv}b. The biggest challenge is data analysis at high degree $l$, where modes of individual degree can no longer be identified and the modal analysis has to be replaced by ``ridge-fitting'' techniques~\citep{Rabello08,Korzennik13,Reiter15}. Better spectral modeling in the intermediate-degree range, where systematic errors of frequency measurements are clearly seen and their origin can be identified, will also bring better confidence to the analysis of lower-degree~\citep{Korzennik05} and higher-frequency~\citep{Rhodes11} domains of the solar-oscillation power spectra.

\subsection{Global Helioseismic Metrology}
Significant efforts are now invested in enhancing the data analysis in global solar seismology. The standard SOHO MDI data analysis pipeline is largely redeveloped, from better accounting for instrumental effects when mapping Dopplergrams to spherical harmonics to more elaborated techniques of frequency fitting \citep{Larson08}, and extended for implementation to SDO HMI data \citep{Larson11}. Below, we address in more detail the prospects of the ``Global Helioseismic Metrology'' project, based on modeling the solar acoustic oscillation in the continuous spectrum.\\

For each pair of target degree $l$ and azimuthal order $m$, the observed power spectrum is approximated by the spectral model defined as \citep{Jefferies06,Vorontsov13b}
\begin{equation}
P_{\rm obs}=\big|L_U+hL_V\big|^2\left\{\left[{A\cos(\varphi-S)\over 1-R^2}\right]^2+B^2\right\},
\quad \tan\varphi={1+R\over 1-R}\tan\theta.
\end{equation}
In this model, $\theta$ is the phase integral of the trapped acoustic wave -- a continuous function of frequency $\omega$ with $\theta(\omega)=\pi n$ at frequencies of acoustic resonances. The energy losses are assumed to be localized in the near-surface layers and described by the surface acoustic reflectivity $R$. The average strength of the stochastic excitation is described by the excitation amplitude $A$. The composite background $B$ is contributed by both the coherent and incoherent components of the solar noise. Parameter $S$ describes line asymmetry, which is governed by the depth and parity of the excitation source and by the coherent component of the solar noise. $L_U$ and $L_V$ are two components of the leakage matrix, which account for vertical and horizontal velocities on the solar surface, $h$ is the ratio of the magnitudes of the two velocity components. The leakage matrix is calculated using a computationally efficient semi-analytic approach described in \citep{Vorontsov05} and extended later to account for non-zero $B$-angle and for the discrete bin sampling implemented in the SOHO MDI ``medium-l'' program. Mode coupling by differential rotation is treated as suggested in \citep{Vorontsov07}.

\begin{figure*}
\centering
\begin{minipage}{1\textwidth}
\begin{center}
\includegraphics[width=0.465\linewidth]{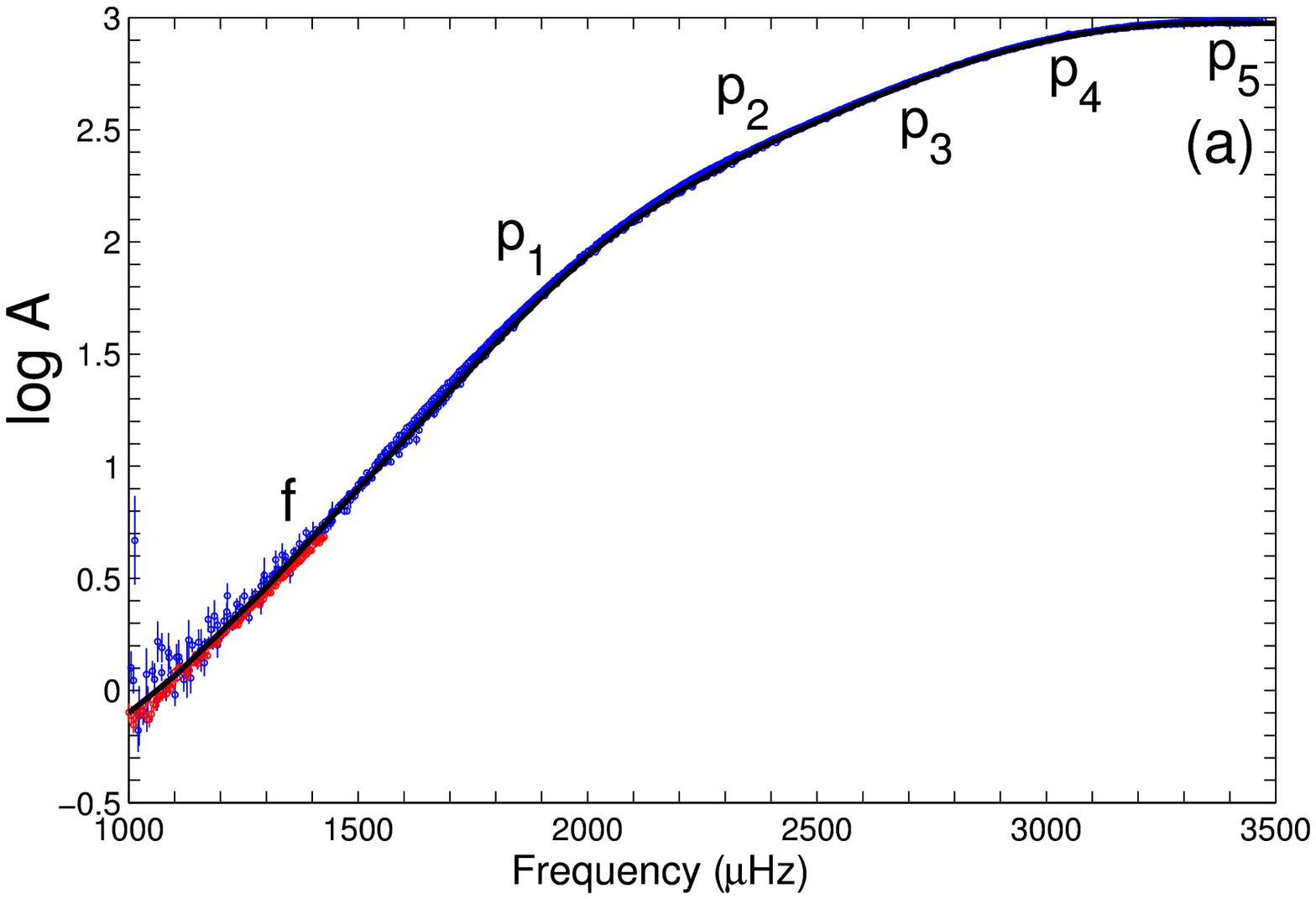}
\hspace{0.1in}%
\includegraphics[width=0.47\linewidth]{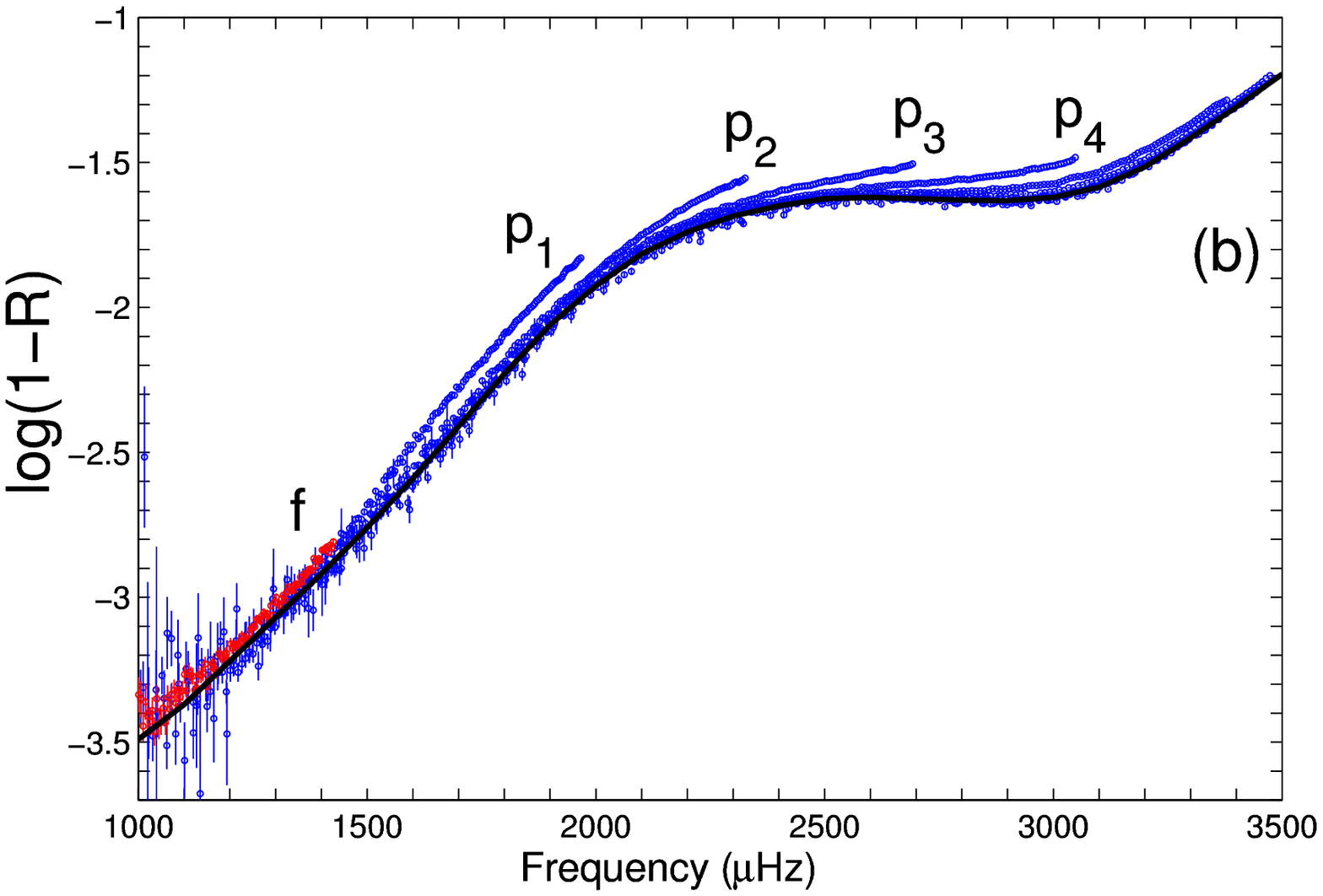}
\vspace{0.1cm}
\end{center}
\end{minipage}
\begin{minipage}{1\linewidth}
\begin{center}
\includegraphics[width=0.465\linewidth]{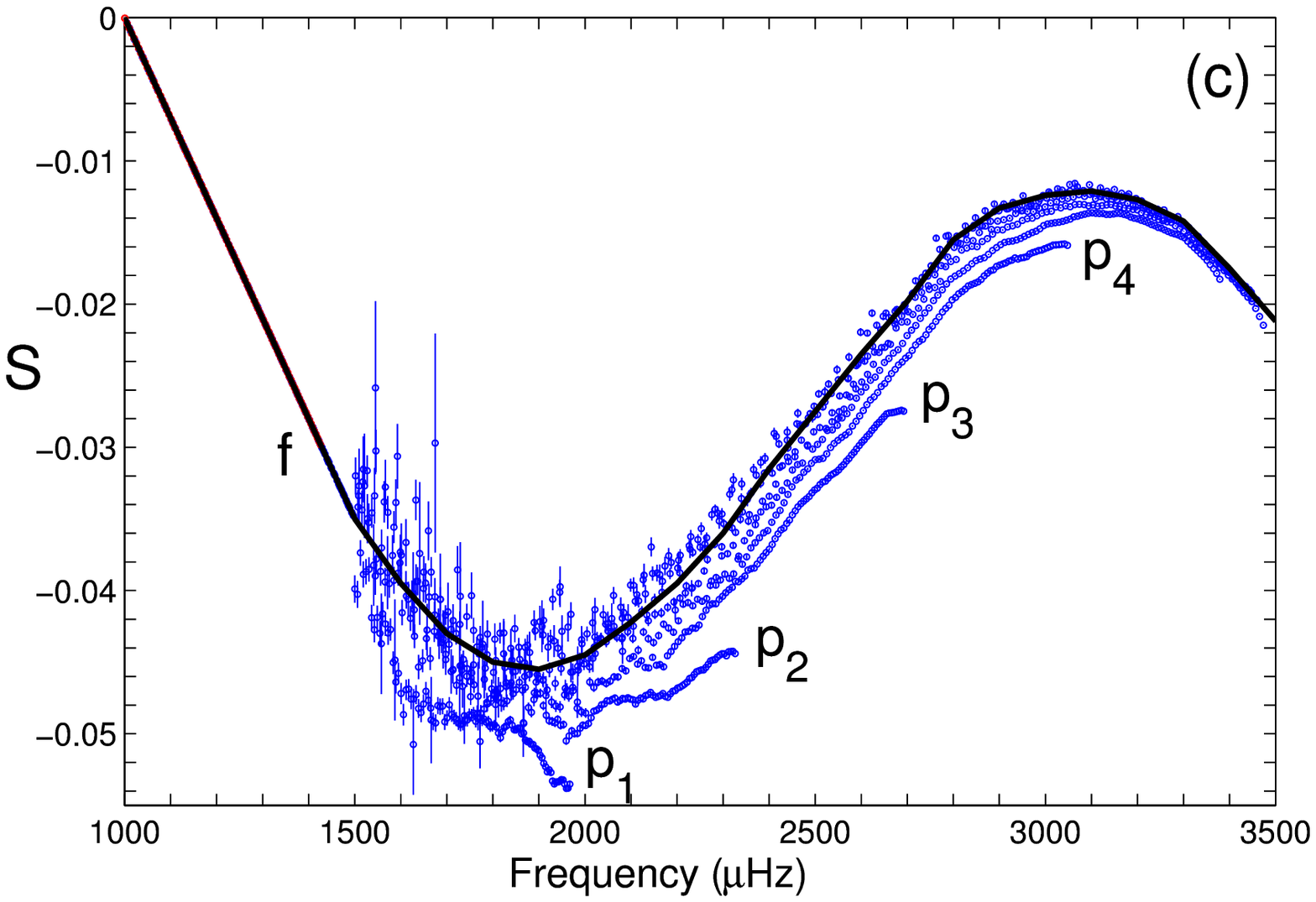}
\hspace{0.1in}%
\includegraphics[width=0.47\linewidth]{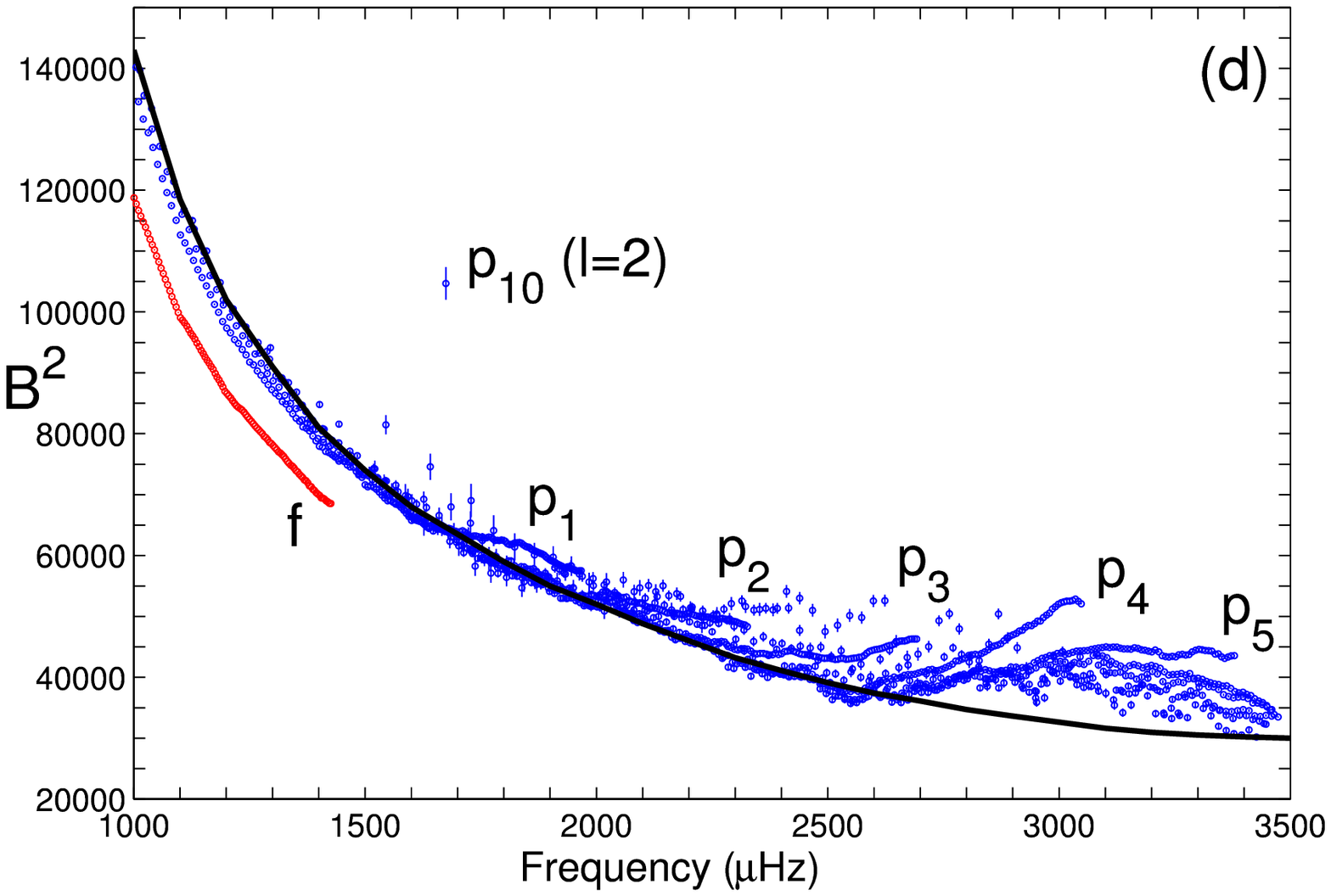}
\end{center}
\end{minipage}
\caption{Excitation amplitude (a), acoustic reflectivity (b), asymmetry parameter (c), and composite background (d) obtained from a 1-year SOHO MDI Doppler-velocity power spectra (adapted from \citealt{Vorontsov13b}). Solid lines show approximations by slowly-varying functions of frequency for lower-degree modes. The results obtained for $f$ modes are shown in red. The asymmetry parameter $S$ is extrapolated by a straight line at frequencies below 1500 $\mu$Hz, where its measurement suffers from uncertainties coming from smaller signal-to-noise ratio and narrow line profiles.}
\label{fig4svv}
\end{figure*}

The parameters of the spectral model, resulted from fitting SOHO MDI power spectra obtained from the first year of observations (at low solar activity) are illustrated by Fig.~\ref{fig4svv}. The results are shown for $p$ modes of radial order $n$ from 1 to 10 and for $f$ modes, in the degree range limited by $l=200$. The maximum-likelihood solution was obtained by an iterative improvement of the spectral parameters $(A, R, S, B)$ and of the resonant frequencies and frequency splittings. The resulted agreement between the model and the data is almost adequate, as can be judged by comparing the $m$-averaged power spectra (Fig.~\ref{fig3svv}). A small systematic inaccuracy in the predicted amplitudes of spatial leaks remains, however. The origin of this mismatch is not yet properly understood; it may be related, in part, with asymmetric distortion of the point-spread function of the MDI instrument \citep{Rabello08}.

An important property of this model, which is behind its diagnostic potential, is that the spectral parameters of individual modes do not depend on the degree and collapse to slowly-varying functions of frequency only when the degree $l$ is not too high (less than about 100). The composite background $B^2$ (Fig.~\ref{fig4svv}d) may look as an exception. However, an accurate measurement of the background at frequencies higher than about 2 mHz is a difficult task, because the background level appears to be significantly smaller than the resonant signals coming from the spatial leaks. As a result, the measurement of $B^2$ can be distorted by small inaccuracies in the leakage matrix. The fitted background is also significantly higher than the average at lowest values of target degree $l$ ($p_{10}$ mode of $l=2$ in Fig.~\ref{fig4svv}d). A possible explanation of this excess is the contribution of the instrumental noise, which can probably be modeled by adding an $l=0$ component to the (otherwise degree-independent) solar $B^2(\omega)$.

Interestingly, the analysis of solar $f$ modes reveal the same values of the excitation amplitude $A$ and ``acoustic reflectivity'' $R$ as solar $p$ modes of similar frequencies (Fig.~\ref{fig4svv}a,b). It indicates that excitation and damping mechanisms do not distinguish between $p$- and $f$ modes, despite the difference in their physical nature (the $f$ modes are incompressible waves). The composite background of $f$ modes appears to be smaller than that of $p$ modes (Fig.~\ref{fig4svv}d). Since at low frequencies the fitted background $B^2$ is dominated by the granulation noise, one possible explanation is that its contribution to the observational power shall be modeled with a smaller (or zero) value of $h$, the ratio of horizontal and vertical velocities. In data processing, a simple theoretical value $h=GM_\odot/(R_\odot^3\omega^2)$ has been used, which corresponds to incompressible motion: this approximation is hardly relevant to the granulation noise.

The major benefit to the $p$-mode data analysis, expected from the global description of spectral parameters, is that the global spectral variables ($A, R, S, B$ as functions of frequency), inferred from the large volume of high-quality intermediate-degree data, can be used in measurements at lower degree $l$. Reducing the number of free parameters in low-degree measurements will bring significant improvement to the accuracy and precision of frequencies (and frequency splittings) of the $p$ modes which penetrate into the deepest solar interior.

\begin{figure*}[t]
\begin{center}
\includegraphics[width=0.8\linewidth]{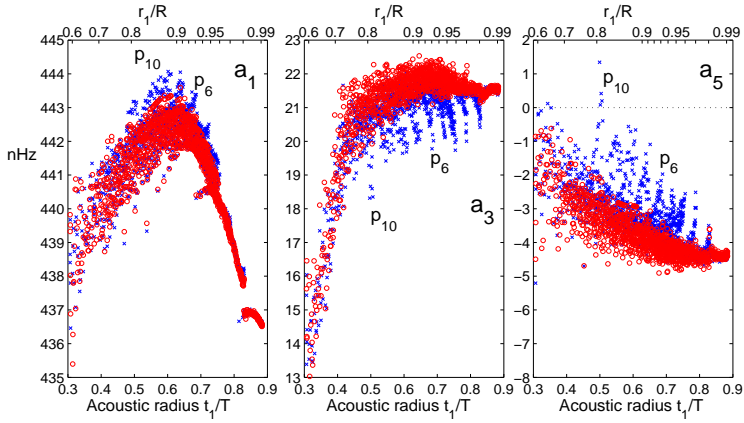}
\end{center}
\caption{First rotational splitting coefficients $a_1, a_3, a_5$, inferred from the first year of SOHO MDI measurements. Blue crosses: published coefficients. Red circles: the result obtained in the ``Global Helioseismic Metrology'' project~\citep{Vorontsov13b}.}
\label{fig5svv}
\end{figure*}

Fig.~\ref{fig5svv} shows the rotational splitting coefficients resulting from the measurement which is described above, in comparison with published splitting coefficients. The horn-like structures, which signify systematic errors (cf Fig.~\ref{fig1svv}) are now eliminated. As indicated by a detailed analysis \citep{Vorontsov09}, the dominant part of the systematic errors came from discarding the effects of mode coupling by differential rotation in the original version of the SOHO MDI data analysis pipeline (which was later improved to include these effects, among others. Systematic errors in centroid frequencies, illustrated by Fig.~\ref{fig2svv}, are due apparently to the temporal variation of the plate-scale error).

\begin{figure*}
\begin{center}
\includegraphics[width=0.45\linewidth]{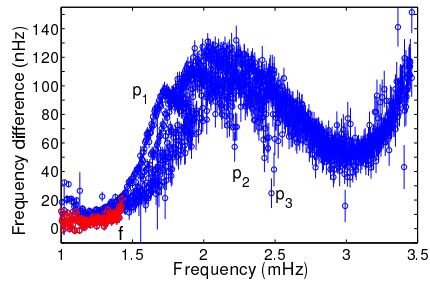}
\end{center}
\caption{Difference between the centroid frequencies measured in the ``Global Helioseismic Metrology'' project from the 1-year SOHO MDI power spectra and their published values~\citep{Vorontsov13b}.}
\label{fig6svv}
\end{figure*}

The differences between the centroid frequencies and their published values are shown in Fig.~\ref{fig6svv}. The major part of the discrepancies is due to the line asymmetries, discarded in the original version of the MDI data analysis pipeline (cf Fig.~\ref{fig4svv}c). Smaller-scale features are due apparently to the combination of the mode-coupling effects with the plate-scale error. The centroid frequencies measured in the ``Global Helioseismic Metrology'' project have been used in a recent study targeted at the seismic diagnostics of the equation of state \citep{Vorontsov13a}. Interestingly, it was found that these frequencies allow the achievement of a significantly better agreement with solar models. This is quite an unusual finding: better accuracy of observational data brings better agreement with theoretical models.\\

The final goal of the project is a new approach to helioseismic inversion, where frequency measurements will be eliminated from the analysis, and the parameters of the rotating solar model will be matched directly with p-mode power spectra. This approach will bring benefits of streamlined regularization, first of all -- by eliminating problems with error correlation and possible mode misidentification in frequency measurements.

\subsection{Global helioseismology from cross-spectral analysis}
In the past, global helioseismology was very successful in measuring the  differential rotation inside the Sun from the splitting of the $p$-mode frequencies. But the meridional flow in the deeper interior is hardly accessible from mode frequencies since the flow is small in amplitude and its influence on the frequencies is only of second order in the flow. Numerical simulations suggest frequency shifts of the order of a few nHz due to the flow~\citep{Roth08,Chatterjee09,Schad11,Vorontsov11}. The perturbation of another characteristic of resonant waves by the meridional flow, the mode eigenfunction, and its potential for helioseismic analyses was early recognized, but largely neglected, likely due the difficulty to measure this kind of perturbation and to discriminate it from systematic instrumental effects in the observations. Below we address the recent advances in using cross-spectral analysis for a global helioseismic measurement of the meridional flow from the perturbation of mode eigenfunctions. 

\subsubsection{Perturbation of mode eigenfunctions by advective mode coupling}
Expressions for the eigenfunctions perturbed by the meridional flow were derived by several authors~\citep{Woodard00,Schad11,Vorontsov11}, who used however partly different assumptions and approximations. The perturbations are considered with respect to a flow-free, purely hydrodynamic reference model of the Sun from which the frequencies and eigenfunctions, $(\omega_{nlm},\mbf{\xi}^{0}_{nlm})$, of the unperturbed $p$-modes are obtained. A flow $\mbf{u}$ leads to an advection of the acoustic waves, which causes a perturbation of eigenfunctions and eigenfrequencies of the $p$-modes with respect to the reference model. If the flow amplitude is small compared to the speed of sound, the perturbed eigenfunctions can be expanded linearly in terms of the unperturbed eigenfunctions: $\mbf{\xi}_{k}=\sum_{k'} c_{kk'}\mbf{\xi}^{0}_{k'}$, where $k$ substitutes the triplet $(n,l,m)$ of the wave numbers of a mode. The perturbations are often considered as a kind of mode coupling and the expansion coefficients $\{c_{kk'}\}$ specify the coupling coefficients which determine the strength of the coupling. \\

Following the general mathematical framework for the perturbation of $p$-modes from \cite{Lavely92}, \cite{Schad11} have shown that the coupling coefficients of a mode $k$ due to the meridional flow are approximated in first order by a linear integral equation, whose integrand specifies the advection kernel,
\begin{equation}
\label{eq:cvals}
c_{kk'}\approx 2\,\I\, \frac{\omega_{k}}{\omega^{2}_{k}-\omega^{2}_{k'}}\int \rho_{0}\, \overline{\mbf{\xi}^{0}_{k'}} \cdot (\mbf{u}\cdot \nabla\mbf{\xi}^{0}_{k})\,d^{3}\mbf{r} \quad \,\mathrm{for}\,\, k'\neq k\,  , 
\end{equation}
where $c_{kk}=1$. For the meridional flow, the coupling coefficients $c_{kk'}$ are purely imaginary. They are expected to be largest for modes of similar frequency and wave numbers $(n,l)$. The meridional flow does not contribute to a self-coupling of modes. As a consequence, the shift of the mode frequencies due to the meridional flow is only of second order in $\mbf{u}$. Assuming azimuthal symmetry for the meridional flow, coupling is only possible between modes of equal harmonic degree $m$ and the coupling coefficient for fixed pairs of $(n,l)$ and $(n',l')$ may be considered as a function of $m$: $c_{kk'}=c_{nl,n'l'}(m)$. This expression can be expanded in terms of polynomials in azimuthal order, where the polynomial order is given by the harmonic degree $s$ of the individual meridional flow components when expanded in spherical harmonics $Y_{s}^{0}$. Different polynomial expansions are suggested. \cite{Schad11} uses a complete set of orthogonal polynomials $\{\mathcal{P}_{l',l}^{s}(m)\}$ that are based on the Clebsch--Gordan coefficients. In the asymptotic case of high degree $l$ and for coupling modes of equal radial order $n'=n$, \cite{Vorontsov11} uses associated Legendre functions which are however not perfectly orthogonal on a discrete grid in $m$. The expression in Equation~\eqref{eq:cvals} defines a linear inversion problem. The polynomial expansion allows to simplify the expression of the coupling coefficients in terms of one dimensional integral equations and to investigate the harmonic components of the flow separately.

\subsubsection{Cross-spectral analysis of solar oscillations}
The oscillating amplitude $\alpha_{k}(t)$ of individual global modes of harmonic degree $l'$ and azimuthal order $m'$ can be extracted by a spherical harmonic transformation (SHT) from sequences of Dopplergrams. Each of these spherical harmonic (SH) coefficients is a weighted sum over the amplitudes of several modes: $o_{l'm'}(t)=\sum_{k}\alpha_{k}(t)\sum_{k''} c_{kk''}L_{l'm',k''}U_{k''}(R)$. The weighting is given by the coupling coefficients $c_{kk''}$, the amplitude of the radial eigenfunction component  $U_{k''}(R)$ of mode $k''$ observed at radius $R$, where the respective absorption line is formed, and the leakage matrix elements $\{L_{k'k''}\}$ of the observing instrument. The leakage origins from the imperfect orthogonality of the spherical harmonic functions when not integrated over the complete solar sphere. The leakage matrix elements are further largely influenced by the apodization mask and the projection of the solar velocity field onto the line-of-sight axis. As a consequence of leakage and mode coupling, modes of a specific target degree $l$ and order $m$ are not perfectly separated from modes of neighboring degree $l'$ and order $m'$ by the SHT. This results in a cross-talk between the SH coefficients and the power of a mode is spread amongst the spectra of spherical harmonic coefficients of neighboring $(l',m')$, where it shows up as sidelobes at the respective mode frequency.\\

\cite{Woodard00} pointed out the sensitivity of the cross-spectrum of the SH coefficients to mode coupling. The cross-spectrum of two SH coefficients $o_{lm}$ and $o_{l'm'}$ is defined by $CS_{lm,l'm'}(\omega)=\langle \tilde{o}^{\ast}_{lm}(\omega)\tilde{o}_{l'm'}(\omega)\rangle$, where $\tilde{o}_{..}(\omega)$ is the Fourier transform of $o_{..}(t)$ and $\langle X \rangle$ denotes the statistical expectation value of the random variable $X$. It is related to the coupling coefficients by
\begin{align}
\label{eq:CSperfect}
CS_{lm,l'm}(\omega)=\sum_{k} S_{\alpha,k}(\omega)\sum_{k',k''}c^{\ast}_{kk'}c_{kk''}L^{\ast}_{lm,k'}L_{l'm,k''}U_{k'}(R)U_{k''}(R)\, ,
\end{align}
where it is assumed that the solar modes are excited independently and $S_{\alpha,k}=\langle |\alpha_{k}(\omega)|^{2}\rangle$ is the auto-spectrum of mode $k$. 

\cite{Schad11} introduced a slightly different quantity to relate the coupling-coefficients with observations: the amplitude ratio $y_{lm,l'm}(\omega_{nlm}):=\tilde{o}_{l'm}(\omega_{nlm})/\tilde{o}_{lm}(\omega_{nlm})$. It measures the relative cross-talk of power of a reference mode with frequency $\omega_{nlm}$ between the SH coefficients due to mode coupling and leakage. Given certain assumptions on the separability of the mode frequencies in the solar oscillation spectrum, which are met only by modes of low and medium harmonic degree $l$, the amplitude ratio is independent of the amplitude $\alpha_{k}$ and can be approximated in first order by~\citep{Schad11}
\begin{equation}
\label{eq:ratio_lineq}
y_{lm,l'm}(\omega_{nlm})\approx\frac{\sum_{k''} c_{kk''}L_{k'k''} U_{k''}(R)}{\sum_{k''} c_{kk''}L_{kk''}U_{k''}(R)} \in \mathbb{C}\, .
\end{equation}
Its expectation value is in leading order determined by the cross-spectrum, since $\langle y_{lm,l'm}(\omega_{nlm})\rangle \sim CS_{lm,l'm}(\omega_{nlm})/S_{nlm}(\omega_{nlm})$. This quantity, denoted as the complex gain, is related to the gain in linear filter theory. The asymptotic statistical distribution of the estimator of the complex gain can be expressed analytically~\citep{Schad13}. 

\subsubsection{Influence of differential rotation}
\begin{figure*}[t]
\includegraphics[width=1\textwidth]{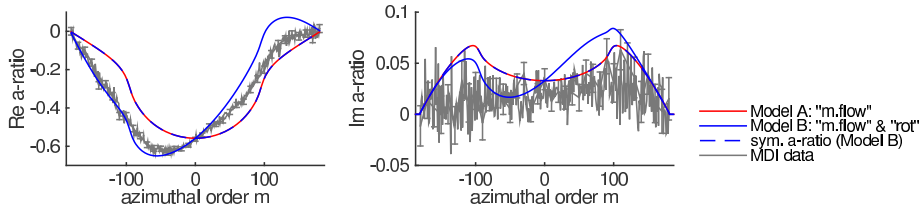}
\caption{Real and imaginary part of amplitude ratios for $(n=1, l=180)$ and $(n'=1,l'=182)$ as a function of azimuthal order $m$~\citep{Schaddiss13}. The amplitude ratios are estimated from SH coefficients from MDI data covering 2004--2010. The ratios are compared to theoretical amplitude ratios computed for simple flow models: Model A) two-cell meridional flow; Model B) two-cell meridional flow and differential rotation adapted to observed rotation rates. The real part essentially exhibits features from leakage and rotation, the latter is responsible for an antisymmetry in azimuthal order. The imaginary part is influenced by leakage, rotation, and meridional flow. Theoretical amplitude ratios for Model B after compensation for rotation by azimuthal symmetrization are shown by a dashed blue line.}
\label{fig:schad1}
\end{figure*}

The toroidal velocity field of differential rotation also leads to a coupling of modes, which cannot be neglected in the eigenfunction perturbation analysis. The perturbation of eigenfunctions due to rotation was investigated by \cite{Vorontsov07,Vorontsov11} for the asymptotic case of large degree $l$. The case of low and medium degree $l$ was investigated by \cite{Schaddiss13}. Formally rotation leads to an additional term to the coupling coefficient: $c_{k'k}=c_{k'k}^{(rot)}+c_{k'k}^{(poloidal)}$. Again, rotation couples only modes of similar frequency and of equal azimuthal order and the coupling coefficients can be expanded by polynomials in $m$ which are equal to the ones used for meridional flow. But in contrast to the coupling coefficients due to the meridional flow, the rotational coupling coefficients $c_{k'k}^{(rot)}$ are real valued and anti-symmetric with respect to the azimuthal order. The different symmetry properties can be exploited to compensate approximately the influence of rotation on the amplitude ratios in analyses for the meridional flow. In this case, the amplitude ratios $y_{lm,l'm}$ are symmetrized with respect to azimuthal order~\citep{Schad13}. Exemplary amplitude ratios estimated from about 6 years of MDI data for $(n=1,l=180)$ and $(n=1,l=182)$, as well as simulated amplitude ratios from numerical forward computations of simple flow profiles, are shown in Fig.~\ref{fig:schad1}. The influence of rotation is clearly visible. Both the real and the imaginary part of the amplitude ratios deviate significantly from azimuthal symmetry if rotation is present. The compensation of the rotational influences by azimuthal symmetrization is also illustrated using simulated amplitude ratios. The symmetrized amplitude ratios matches very well to the amplitude ratios obtained for a velocity field without solar rotation.\\

\subsubsection{Application to data: Measurement of the meridional flow}
First measurements of the meridional flow from analysis of the perturbation of $p$-mode eigenfunctions were given by~\cite{Schad12,Schad13,Woodard12}. 

\begin{figure}[h]
\vspace{0.5cm}
\begin{center}
\begin{minipage}{0.49\textwidth}
\centering
  \includegraphics[width=0.9\textwidth]{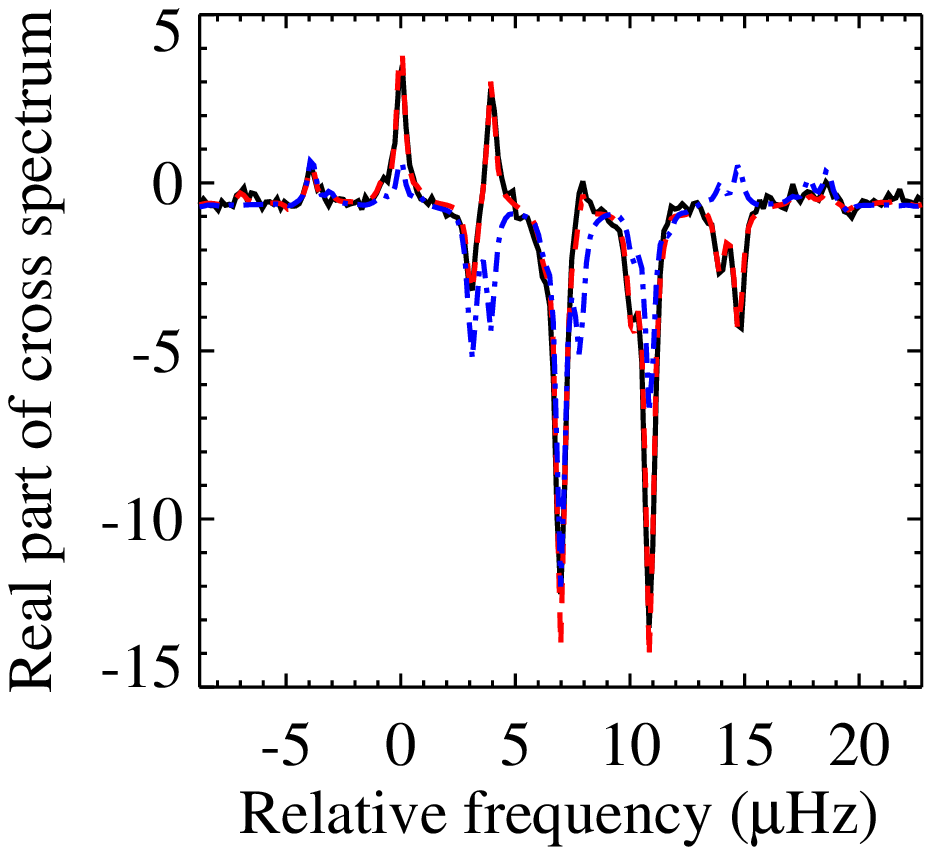}
\end{minipage}
\begin{minipage}{0.49\textwidth}
\centering
  \includegraphics[width=0.9\textwidth]{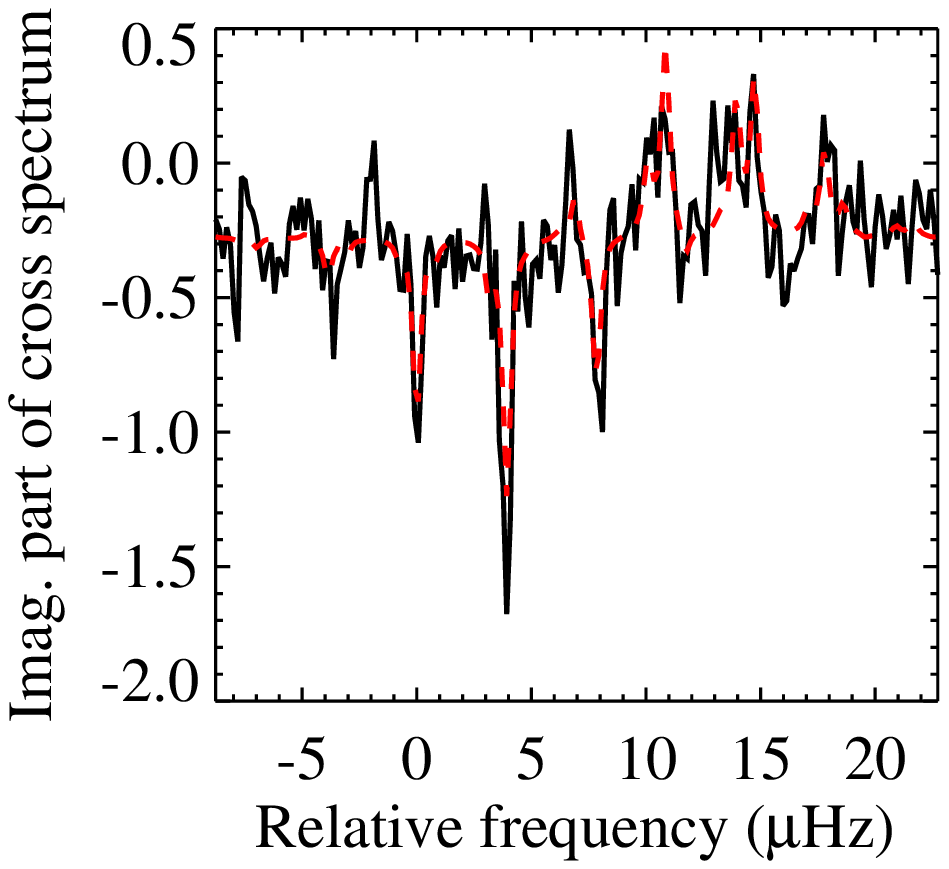}
\end{minipage}
\end{center}
\caption{
De-rotated cross-spectra for $l=200$ and $l'=202$ averaged over positive order $m$ as a function of $\nu$ relative to the frequency of the f-mode multiplet of degree $l$ (Figures adapted from~\cite{Woodard12}). Left: Real part. Right: Imaginary part. The black curves depict the cross-spectrum estimated from 500 days of HMI data. The blue dash-dotted curve is the fitted cross-spectrum when ignoring the effect of flows on mode eigenfunctions. The red dashed curves are fitted cross-spectra including the effect of differential rotation and meridional flow.}
\label{fig:schad2}
\end{figure}

\cite{Woodard12} fitted a model to cross-spectra estimated from HMI data to measure the horizontal component of the meridional flow. The model for the cross-spectrum incorporates instrumental leakage, differential rotation, and the solar background. They analyzed HMI data with a length of 500 days and investigated couplings between modes of the same radial order $n$ for the harmonic degrees $20\leq l \leq 300$. Cross-spectra estimated from the HMI data and averaged over azimuthal order as well as cross-spectra after fitting different cross-spectral models to the data are depicted in Fig.~\ref{fig:schad2}. The comparison illustrates the improvement of the cross-spectral model by inclusion of flow-dependent eigenfunction perturbations~\citep{Woodard12}. 

\begin{figure}[t]
\begin{minipage}{1\textwidth}
\begin{minipage}{0.5\textwidth}
\begin{center}
  \includegraphics[width=1\textwidth]{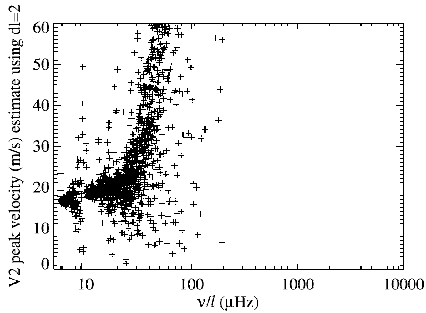}
\end{center}
\end{minipage}
\begin{minipage}{0.5\linewidth}
\vspace{-1.2cm}
\captionof{figure}[]{Surface peak velocities for the horizontal flow component $V2$ estimated from a cross-spectral analysis of HMI data as a function of $\nu/L$ as found by \cite{Woodard12}. The steep increase of the peak velocity with depth toward speeds $>30$\,m/s is suspected to be due to systematic errors~\citep{Woodard12}.}
\label{fig:schad3}
\end{minipage}
\end{minipage}
\end{figure}

The estimated peak velocities of the horizontal flow component with harmonic degree 2 ($V2$) are shown in Fig.~\ref{fig:schad3} as a function of $\nu/L$ that is related to the lower turning point of the acoustic waves~\citep{Woodard12}. 
Near the surface, the horizontal peak velocities are of the order of about 20\,m/s as expected from local helioseismic measurements. Toward the interior, the velocities exhibit an unexpected large increase and the authors assume that their measurements are likely affected by a systematic effect~\citep{Woodard12}.\\ 

\begin{figure}[t]
\begin{center}
\includegraphics[width=0.7\textwidth]{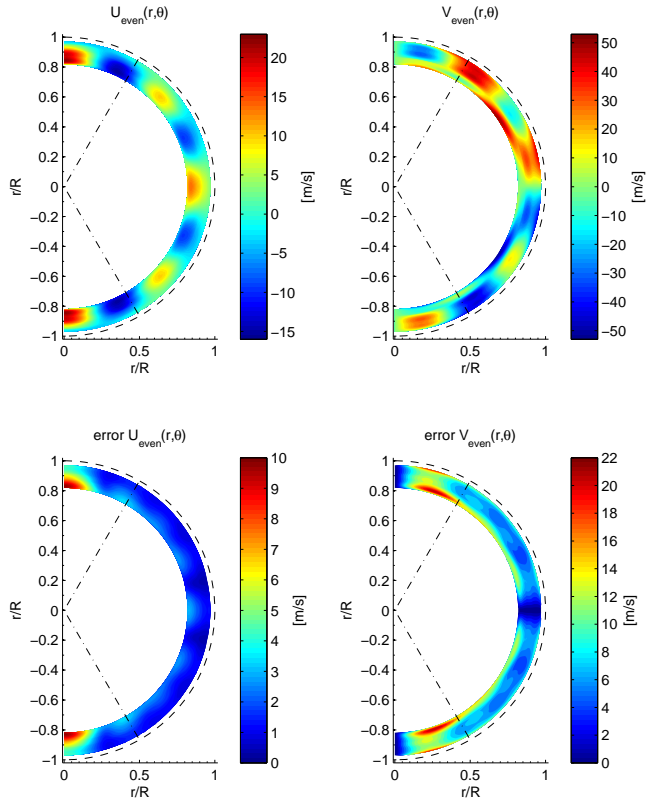} 
\end{center}
\caption{Top row: The meridional flow estimated from MDI data covering 2004--2010 as a function of radius $r/R$ and latitude~\citep{Schad13}. The profile is obtained from a composite of flow components of the harmonic degrees $s=2,4,6,8$. A dashed line indicates the solar surface. Left: Radial flow component. Positive (negative) values of $U$ refer to an outward (inward) directed radial flow. Right: Horizontal flow component. Positive (negative) values of $V$ refer to a northward (southward) directed horizontal flow. Bottom row: $1\sigma$-error on the respective flow estimates.}
\label{fig:schad4}
\end{figure}

\cite{Schad12,Schad13} set up a global helioseismic estimation scheme for the meridional flow based on the analysis of amplitude ratios of coupling modes. It takes into account leakage and the influence of rotation is compensated by symmetrization of amplitude ratios. They applied their method to MDI data covering 2004--2010 for modes with $0\leq l\leq 200$ and investigated couplings between modes of different radial order and for flow components of harmonic degree $s=1,\dots,10$. They found two flow components of harmonic degree $s=2$ and $s=8$, which differ significantly from zero. Individual components, like the $s=8$ flow component were measured deep down to 0.5\,R. The radial component of the flow $U(r,\theta)=\sum_{s}u_{s}(r)Y_{s}^{0}(\theta,\phi)$ is estimated from a composite of the individual flow components. The horizontal flow component $V(r,\theta)=\sum_{s}v_{s}(r)\partial_{\theta}Y_{s}^{0}(\theta,\phi)$ is reconstructed from the radial components $\{u_{s}\}$ assuming mass conservation. The composite of flow components of even degree $s=2,\dots,8$ are depicted in Fig.~\ref{fig:schad4} as a function of radius and latitude. The meridional flow exhibits a multi-cellular pattern over latitude and depth. Near the solar surface, the horizontal flow is consistent with subsurface flow measurements from ring-diagram analyses indicating a poleward directed flow on each hemisphere with a small-scale latitudinal modulation and a speed of about 20\,m/s at mid-latitudes. Their findings substantiate the assumption that the flow is confined between the tachocline region and the solar surface.

\subsection{Discussion}
The systematic features of the estimates of global mode parameters considered in Sec.~\ref{sec:global} were found from analyses of medium-$l$ MDI data and can be reduced by the approach used by the global metrology project. Parameters estimated from data from the same instrument but of different preprocessing or from data from other instruments, e.g., HMI or GONG, may show other systematic features. But independent of the instrument, one cannot avoid for example leakage of mode power due to the observational restrictions or mode coupling dominated by rotation.

The development of global analysis methods based on mode eigenfunction perturbations and the analysis of cross-spectra of spherical harmonic Doppler velocity coefficients is a promising emerging field for inferences on the meridional flow in the deep solar interior. In this approach, care must be taken of mode eigenfunction perturbations from other sources, for example from differential rotation. Vice versa, analyses of mode eigenfunction perturbations may also be used to investigate rotation. Next to the solar subsurface, the meridional flow measured by~\cite{Schad13} seems to be consistent with flow measurements from a ring-diagram analysis. But, the recent measurement of the meridional flow from a deep focusing time-distance helioseismic analysis of HMI data by \cite{Zhao13} seems to disagree with their findings. However, the investigated HMI data cover a later observation time period. Further investigations are necessary to test the reliability of these results.

The success of each of the global helioseismic methods considered here strongly depends on the accurate knowledge of the leakage matrix. The inferences from global helioseismic analyses suffer from systematic effects if the leakage matrix is inaccurate or the models used to estimate the helioseismic parameters from spectra or cross-spectra are misspecified. For example, a comparison study of splitting coefficients and rotation profiles obtained from medium-$l$ and full-disk MDI data indicate that the there used leakage matrix used for the medium-$l$ MDI data probably does not perfectly account for the apodization function and that might be responsible for a spurious polar jet in the solar rotation profile from the medium-$l$ MDI data~\citep{Larson09}. 

The results obtained so far clearly illustrate the necessity of putting efforts into enhancing the models of the solar oscillation spectrum and of improving the accuracy of the leakage matrix of the respective instruments. A difficulty in the computation of the leakage matrix comes along with taking into account systematics that change over time, like the $B$-angle, or from not properly known instrument characteristics, like the plate-scale error or the point spread function.

\section{Data assimilation in solar physics}
Understanding and predicting the solar activity cycle poses one of the main problems in solar physics and comes along with questions about the timing, amplitude, and shape of the currently evolving and following cycle. Convection, rotation and the mean meridional flow are thought to be key ingredients that drive the generation and evolution of the solar magnetic field. In the previous sections it was shown that new helioseismic techniques could provide estimates of some of these ingredients. We will see in this section how observations and helioseismic measurements could be combined with physical models to improve our knowledge of the solar activity cycle.

We have entered an era where extremely large amounts of data are available concerning the Sun, thanks to the high-resolution observations of satellites like Hinode or more recently SDO. At the same time, considerable progress has been made on the multi-dimensional numerical simulations of highly non-linear physical processes interacting in the solar interior and in its atmosphere. Although still far from realistic values of the parameters, several impressive local and global computations now seem to reach high levels of turbulence and capture relevant physical processes occurring in our star. Various 3D MHD codes are used for that purpose, among which the ASH code \cite[e.g.][]{Miesch08, Brun11, Nelson13, Alvan14}, the PENCIL code \cite[e.g.][]{Kapyla12, Warnecke14}, the MURaM code \cite[e.g.][]{Cheung10, Rempel14} or the STAGGER code \cite[e.g.][]{Stein12}. This improvement is intimately linked to the recent fast development of high performance computing, several computers in the world now reaching the performance of tens of PetaFlops. Finally, we live today in a technological society in which strong solar flares, CMEs or any violent events linked to solar activity could cause significant damage to satellites, air traffic or telecommunication networks. That is why a solar cycle panel, whose role is to produce predictions of solar activity, was created in 1997 and has provided us with estimates of the sunspot number for Cycle 23 and current Cycle 24. 

As Cycle 24 progressed, it became clearer and clearer that it would be a weak cycle. A quantitative estimate of the cycle strength can be given by the Wolf number, defined as $R=k (10 g+s)$, with $g$ the number of sunspot groups, $s$ the total number of individual sunspots in all groups and $k$ a variable scaling factor that accounts for instruments or observation conditions. We now know that for Cycle 24, the monthly smoothed Wolf number reached a peak of about 82 in April 2014. This will probably become the official maximum. This second peak surpassed the level of the first peak (about 67 in February 2012). Many cycles are double peaked but this is the first in which the second peak in sunspot number was larger than the first. These features make Cycle 24 the weakest cycle since Cycle 14, which peaked in 1906. Among the predictions, less than $25\%$ had anticipated such a small number. It thus still seems extremely difficult to provide reliable predictions of the long-term solar activity with the techniques that have been used so far (mostly relying on geomagnetic precursors or other statistical estimates -- see \citealt{Hathaway09} for a review on the subject). In meteorology, data assimilation which cleverly combines observational data and numerical models has been used for decades and now routinely in weather-forecasting. Considering the high-quality observations at our disposal, the recent progress in numerical simulations and the necessity to produce predictions of the amplitude, timing and shape of the next solar cycle, it seems rather reasonable to try to apply data assimilation to solar physics \citep{Brun07}.

\subsection{First attempts to introduce data into models}
There have been first attempts to connect models and data, not exactly through data assimilation but rather by driving models with a time series of well-chosen data. This procedure has been implemented for flux-transport dynamo models and for photospheric and heliospheric magnetic field evolutions. In their mean-field dynamo model, \citet{Dikpati06} have introduced a surface source of magnetic field that depended upon the sunspot areas observed since 1874, when \citet{Choudhuri07} chose the surface field at minimum to be their driving observations. Both models produced good agreement with previous observed cycles but differed completely on the predictions for Cycle 24. One of the reasons was that one model (Dikpati's) was dominated by the advection process while the other (Choudhuri's) was dominated by diffusion, changing drastically the characteristics of the memory of the system, as was shown by \cite{Yeates08}. It should be noted that both predictions happen to be quite far from reality, either on the timing of the cycle, or its amplitude. Data-driven models for the solar atmosphere now also tend to develop. A first attempt was made by \citet{Schrijver03} when they introduced SOHO/MDI magnetograms into a flux-dispersal model to see the influence on the coronal reconfigurations. More recently, \citet{Cheung12} simulated the evolution of the active region coronal field driven by temporal sequences of photospheric magnetograms from SDO/HMI. Under certain conditions, they found that data-driven simulations could produce flux ropes that were ejected from the modeled active region due to loss of equilibrium, possibly showing a way to predict violent events through simulations.

In the future, we would like to make observations and models really work together and feed the simulations with data so that the model improves itself and allows to make reliable predictions. To do so, data assimilation techniques are exactly what we are looking for (see \citealt*{Kalnay03} for a general introduction related to atmospheric sciences and \citealt*{Fournier10} for applications to geophysics). Indeed, they consist in combining observational data and numerical models to produce what is called an analysis: the new information provided by the observational data is taken into account in order to advance in time the ``background'' state that the numerical code has predicted. The increment is obtained by taking the difference, or innovation, 
between the observational data and the observation operator. More specifically, let ${\bf x^b}$ be the background vector state characterizing 
the current state of the model, $H$ the observational operator and ${\bf y^o}$ the observational data to be assimilated in the model, 
then one can show that the analysis ${\bf x^a}$ is:

\begin{equation}
{\bf x^a}={\bf x^b} + W({\bf y^o}-H({\bf x^b})),
\label{eq1}
\end{equation}
where $W$ represents weights whose exact determination will differ from one assimilation technique to another. Two approaches and their applications to solar physics will now be briefly presented and discussed.

\subsection{Sequential assimilation and application to solar physics}
In the sequential assimilation technique, the background state is advanced in time thanks to the numerical model and corrected (i.e. an analysis is performed) each time an observation is available. The analysis is performed using Equation~\ref{eq1}, where $W$ is given by the Kalman gain matrix which is a combination of the covariance matrices of the forecast and of the observational errors.

\begin{figure}[h!]
\centering
\includegraphics[width=8cm]{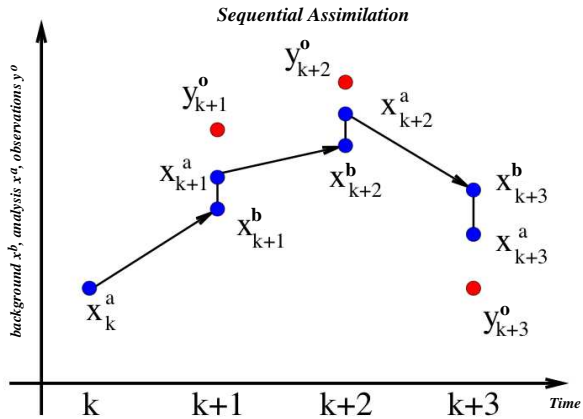}
\caption{Schematic representation of the sequential data assimilation method used in weather forecast \citep[adapted from][]{Bocquet15}. The background state ($x^b$) is updated every time an observation is available and the model evolves the state until the next step (following the arrows), at which observational data ($y^o$) are again assimilated to produce the analysis ($x^a$). }
\label{fig1}
\end{figure}

The analysis at time $k$ thus provides the new state vector $x_k^a$ from which the forecast is calculated by running the numerical model. This forecast step thus produces the new background $x_{k+1}^{b}$ which in turn will be corrected by the observation at time $k+1$. This sequential process is illustrated in Fig.~\ref{fig1}. This technique thus propagates information forward in time and also gives an estimate of the forecast errors and of their evolution.

As far as predictions of future solar activity is concerned, sequential assimilation has been implemented for the first time by \citet{Kitiashvili08} in a 1D mean-field dynamo model evolving jointly the three components of the magnetic field and a measure of the magnetic helicity. The observations used were the annually smoothed Wolf sunspot number for the period 1857-2007. To derive the observational operator $H$, the following relationship between the Wolf number and the toroidal (in the azimuthal direction) magnetic field was used: $\vert B_{\phi}\vert \propto R^{2/3}$. The model used was a classical $\alpha\Omega$ dynamo model in which the toroidal field owes its origin to the differential rotation shearing the poloidal (in the meridian plane) field lines (the $\Omega$-effect) and where the poloidal field is due to helical turbulence within the solar convection zone acting on the toroidal field (the $\alpha$-effect). This model contains a number of simplifications (see \citealt*{Brun13} and \citealt*{Charbonneau10}), most notably a rather crude parameterization of the effects of turbulence on the large-scale magnetic field but has the advantage of producing a cyclic variation of ${\bf B}$ and an exponential growth of the magnetic energy (saturated here by the back-reaction of the magnetic field on the $\alpha$-effect).

\begin{figure}[h!]
\centering
\includegraphics[width=8cm]{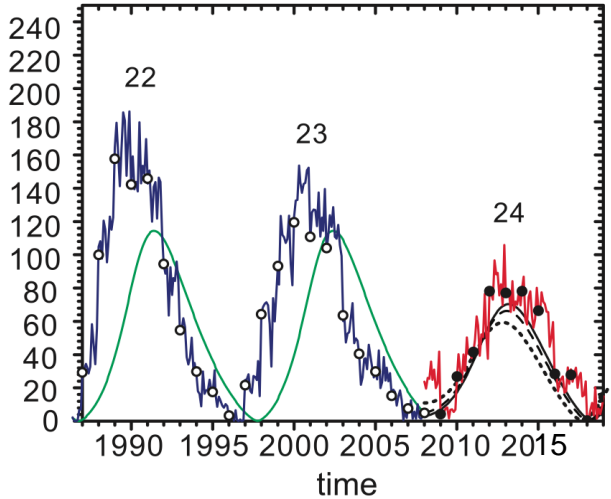}
\caption{Prediction of the sunspot number for Cycle 24 from \citet{Kitiashvili08}. The green curve shows the model reference solution. The blue curve shows the best estimate of the sunspot number using the observational data (open circles) and the model outputs (black dots), for the previous cycles. The black curve the model solution according to the initial conditions of the last measurement. The red curve shows the prediction results. The model solution is shown for three different estimates of the sunspot number for 2008: 3 (black curve), 5 (dashed curve), and 10 (dotted curve).}
\label{fig2}
\end{figure}

Fig.~$\ref{fig2}$ shows the result of their forecasting step for Cycle 24, after they applied their sequential assimilation to get the analysis at the time they wrote the paper (2008). Their predictions were shown for three different estimates of the Wolf number in 2008 since they did not have yet the real data. What we see on this figure is actually quite a good agreement with what Cycle 24 looks like now, reaching its maximum of around 80 in 2013 (a bit early then). However, we should keep in mind that the model was here very simple and the relationship between the observational data and the outputs of the model rather uncertain, leading them to state in the conclusion that additional observations like the latitudinal distribution of sunspots should probably also be taken into account to make progress in the subject. 

First steps towards using a more complete model including a large-scale meridional circulation were undertaken recently by \citet{Dikpati12} and \citet{Dikpati14}. Their idea is to use a 2D dynamo model in which the poloidal field is generated by the decay of active regions emerging at the solar surface \citep{Babcock61,Leighton69} and not by small-scale turbulence in the convection zone like in classical $\alpha\Omega$ dynamo models. The data they are planning to use is similar to what \citet{Kitiashvili08} chose, namely the monthly smoothed sunspot number data from the Royal Observatory of Belgium but they also intend to make efficient use of data concerning the solar meridional circulation. Their first step was thus to determine the response time of the whole system to perturbations of the meridional flow, which is not very well measured below the first few Mm of the Sun and which is known to produce large changes in the timing and possibly the shape of the magnetic cycles in those types of flux-transport dynamo models \citep{Dikpati99, Jouve07}. Dikpati \& Anderson found that in the advection-dominated regimes they considered, the models time of peak response to a change in flow speed was rather short compared to the full circulation time and thus that a modification of the amplitude of the flow would quickly show large changes in the evolution of the magnetic field. In a recent subsequent paper \citep{Dikpati14}, they applied a sequential data assimilation technique to reconstruct the meridional flow speed at the solar surface (fixing the meridional flow profile to one large circulation cell per hemisphere) from synthetic observations of the magnetic field. The synthetic observations were produced by running the model with a fixed meridional flow profile and speed and then noised to mimic observational errors.  They found that the best reconstruction of meridional flow-speed could be obtained when 10 or more observations were used with an up-dating time of 15 days and an observational error of less than $10\%$.

Another relevant quantity to assess when dealing with forecasting in such dynamical systems is what \citet{Lhuillier11} call the forecast horizon or the time interval over which reliable predictions can be achieved. They made a detailed analysis of the growth rate of perturbations applied to the magnetic, velocity or temperature field for geodynamo simulations and found that the limit of predictability would be a combination of this growth time (estimated to be of about 30~yr), of all types of errors affecting the initial conditions and of the limited numerical resolution. Recently, \citet{Sanchez14} performed the same kind of analysis for a flux-transport mean-field dynamo model, similar to the one considered by \citet{Dikpati14}. They measured the rate associated with the exponential growth of an initial perturbation of the model trajectory, and found a characteristic $e$-folding time of 2.76 solar cycle durations. These results are quite promising for possible future predictions of solar activity. However, thorough studies of the sensitivity to all model parameters and on the predictability skills of these models will have to be undertaken before being able to apply more complete sequential data assimilation and to rely on predictions coming from simplified solar dynamo models.

\subsection{Variational assimilation and applications to solar physics}
\begin{figure}[h!]
\centering
\includegraphics[width=8cm]{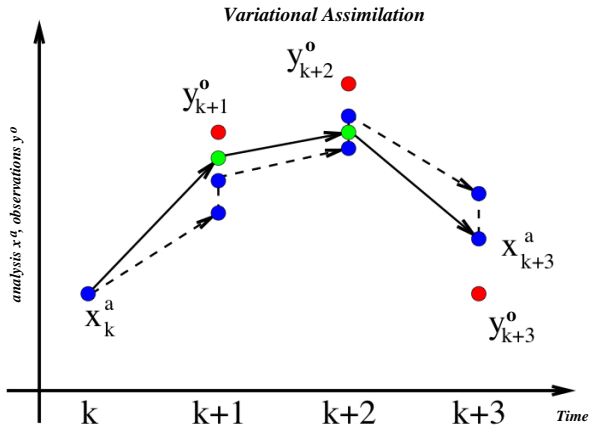}
\caption{4D variational method and comparison with sequential assimilation~\citep[adapted from][]{Bocquet15}. In the 4D variational method, within a time interval the model and the observations are taken into account in the cost function that needs to be ''minimized''. The minimization of this cost function results in a best trajectory (plain arrows) across the observations.}
\label{fig3}
\end{figure}

As opposed to sequential assimilation, the variational technique consists in adjusting the trajectory of the model through observations over a significant time interval. This is illustrated in Fig.~\ref{fig3}. To be more formal, the analysis $x^{a}$ is found here by minimizing a cost function $J$, defined over the entire time window where observations are available. This cost (or objective) function is the sum of two terms. The first one measures the distance between the outputs of the model and the observations and the other one measures the distance to an a priori estimate of the background state, if there is any. This second term is called the background term, it can be used to provide the system with information about regularity of the solutions we are looking for, approximations that our flows need to satisfy or typical values we expect at certain points in the domain. After the analysis is found, the forecast step is similar to the one performed in the sequential technique, it is calculated by applying the numerical model to the analysis which has just been determined. This technique thus propagates information both forward and backward in time since the present state is estimated using the past and future observations available over the entire time window. This is important if we wish to reassimilate past data or in other words propagate backward in time the current quality of observational data. This advantage of variational assimilation might be of great use for solar physics for example if we wish to have a better insight on the state of the system at periods where observations were missing because of a lack of instruments or of a lack of surface events, like in the periods of grand minima. However, as stated before, we should of course keep in mind that the quality of the forecast will be strongly limited by the range of predictability in such dynamical systems \citep{Lhuillier11}.

The main drawback however of using variational assimilation is that it requires the development of an adjoint model which will be used to provide the gradient of the cost function with respect to all input variables. Both the values of the function itself and of its gradient are then combined in a minimization algorithm to produce the analysis. Simple recipes can be used to write the adjoint code of a numerical program by hand \cite[e.g.][]{Talagrand91, Giering98} but for very large codes, it can be tempting to resort to automatic differentiation (AD) algorithms. AD is becoming a very efficient and powerful tool used to produce the adjoint code of general circulation models in meteorology (among many others, we can quote the on-line tool \emph{TAPENADE}, developed by the TROPICS team in INRIA Sophia Antipolis, France).\\

Variational assimilation (or 3D/4D-VAR) was used recently in astrophysics for the problem of 2D stratified convection \citep{Svedin13} and in the context of solar physics for two main applications as of today: models of solar flares and dynamo models. In both studies, only synthetic observations were used, i.e. produced by a model and not by nature. Nevertheless, those are first steps towards understanding how variational assimilation may help us understand the current state of our Sun and hopefully predict its magnetic activity. In \citeyear{Belanger07}, \citeauthor{Belanger07} used a phenomenological model, called the avalanche model, which, although a priori far removed from the physics of magnetic reconnection and magnetohydrodynamical evolution of coronal structures, nonetheless reproduces quite well the observed statistical distribution of flare characteristics. This model is the continuous analogous to one of a sandpile where grains are dropped one by one until the pile reaches an equilibrium conical shape. Addition of more grains will then sometimes produce small to large avalanches or may have no consequences at all, leading to a strongly intermittent unloading while the loading remains slow and gradual. The cost function which is minimized here is the misfit between the outputs of the model and the synthetic observations (produced with a known set of parameters), in terms of an amount of energy released, averaged in time. They show that, despite the unpredictable (and unobservable) stochastic nature of the driving/triggering mechanism within the avalanche model, 4D-VAR succeeds in producing optimal initial conditions that reproduce adequately the time series of energy released by avalanches and flares. More recent works about the predictability of solar flares with the avalanche model have however shown that only a modified version of it, purely deterministically driven, could produce reliable predictions for large eruptive events \citep{Strugarek14}.

\begin{figure}[th]
\centering
\includegraphics[width=13cm]{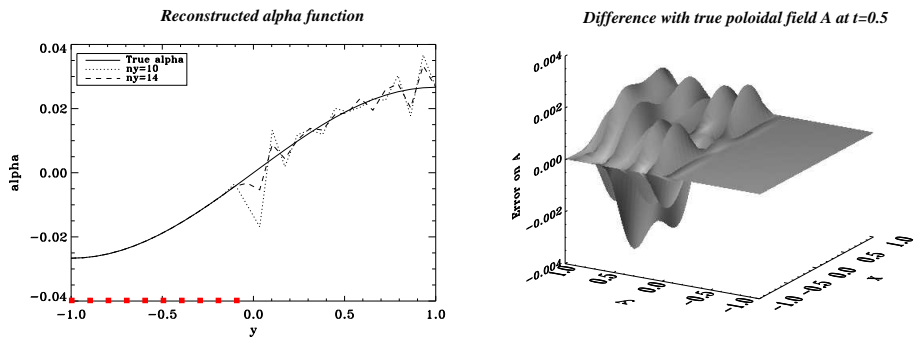}
\caption{Adapted from \cite{Jouve11}. Left panel: $\alpha$ function reconstructed after assimilation of data in the Southern hemisphere only, with 2 different sets of observations, superimposed with the true state. The red dots indicate the locations of the observations in the case were we have 10 observations in latitude ($ny=10$). Right panel: difference between the poloidal potential produced by the reconstructed $\alpha$ and the true state at t=0.5 (in the middle of the time interval).}
\label{fig4}
\end{figure}

Variational data assimilation was also applied in the context of solar dynamo models. Since the idea is to produce predictions of future solar activity, trying to use a technique which is now used routinely in weather forecasting on Earth sounds like a reasonable way to go. In \citet{Jouve11}, a variational data assimilation technique was developed, using a 2D mean-field $\alpha\Omega$ dynamo model. As we said before, synthetic data were used, consisting of outputs from the model produced by a particular set of parameters and more specifically a particular choice of $\alpha$ as a function of latitude (designated by the coordinate $y$ here). After the observations were produced, a cost function was chosen to be the misfit between the model toroidal field and the observations of the same quantity. The result of the minimization was then supposed to reproduce the $\alpha$ function used to calculate the observations. The authors performed an analysis of the dependence of the quality of this recovery as a function of the number and location of observations. In Fig.~\ref{fig4}, an example of such a calculation is shown, where data were only present in one hemisphere (from $y=-1$ to $y=0$). The recovery of the $\alpha$ function was very good in the hemisphere where data were present but the difference with the real solution (called the true state) was also found to be reduced in the other hemisphere when more observations were assimilated. This indicates that the cost function (and thus the values of the magnetic field) in one hemisphere was sensitive to the $\alpha$-effect in the other hemisphere, which is the kind of insight that can be gained from variational data assimilation techniques. 

One of the challenges in solar physics that can be tackled under the angle of data assimilation is to get a better knowledge of the meridional flow and its role in the dynamo process. Hung, Jouve, Brun, Fournier \& Talagrand, 2015 (in preparation) are now developing a variational data assimilation technique applied to a 2D spherical flux-transport dynamo model. This will allow consideration of various possible observations beyond the basic averaged sunspot number. For the magnetic field, one could use the amount of poloidal field at the poles, the timing of its reversal (that we can get from Hinode or SDO for example), the components of the multipolar expansion of the magnetic field (accessible of course on the Sun, \citealt*{DeRosa12}, but also now for other stars through spectropolarimetry, see \citealt*{Petit08}). For the velocity field, use could be made of the amplitude and profile of the meridional flow (both at the surface and deeper down thanks to new helioseismic techniques, as discussed in this paper) and possibly of torsional oscillations \citep{Vorontsov02b, Spruit03}. By applying data assimilation with those various sources of observations and an a priori knowledge of part of the meridional flow, the first results of this study indicate that it is possible to reconstruct not only its amplitude as in the work of \citet{Dikpati14} but also its profile in the solar interior and its structure close to the base of the convection zone, which are of prime interest for dynamo modelers.

\subsection{Perspectives for data assimilation in solar physics}
The applications presented here are of course very preliminary in terms of models and observations. Firstly, they do not yet use real data but were only tested on synthetic observations or on very smoothed proxies. Secondly, the models are extremely simple compared to the most up-to-date 3D models evolving the full set of magneto-hydrodynamics equations in spherical geometry. However, and that is maybe what constitutes a big difference with meteorology, we have limited access to observations within the solar interior (with the notable exception of what is learned thanks to helioseismology and particularly new global and local techniques, as is shown in this paper) and moreover we do not have yet at our disposal a self-consistent 3D dynamo model reproducing the main characteristics of the large-scale solar magnetic field. We thus have to move step by step towards this goal by first considering surface observations assimilated in simplified solar models. 

In this context, it could be useful to work on ensemble forecasting, similar to what is used today in weather predictions on Earth. In the case of weather forecasting, the idea is to make several predictions starting with slightly different initial conditions and to take an average of those predictions to get the actual forecast. This has mainly three goals: improving the forecast thanks to the averaging, providing an indication of the reliability of the prediction and giving a quantitative way to assess the quality of each individual forecast. In solar physics, we could think of forming an ensemble not by perturbing the initial conditions for the same model but rather by considering different models, where the key physical processes may have more or less impact on the evolution of the magnetic field. Each model would then provide its own forecast. Eventually, the averaging process would give us a way to distinguish between those various models and would indicate which characteristics of the cycle we are the most likely to correctly anticipate. That is probably a future development to be considered on the way towards better and more reliable long-term predictions of solar activity.

\section{Summary}
In the first part of this paper we reviewed some recent developments of local and global helioseismic data analysis methods. For the local helioseismic analysis we focused on the time-distance method as used for the estimation of velocity fields in the solar subsurface, e.g., supergranulation. Such analyses of rather rapidly evolving processes benefit from sophisticated averaging schemes and filtering methods that help to increase the signal-to-noise ratio and to retrieve signatures from travel--time shifts.

Regarding global helioseismology, we considered two advances, the metrology project and the mode eigenfunction perturbation analysis for inferences of the meridional flow. Global helioseismic inferences rely on the accuracy and reliability of estimated global seismic mode parameters, like the mode frequencies. The global helioseismic metrology project aims to improve accuracy and reliability of the parameter estimates by better incorporation of systematic influences in the parameter estimation scheme. The systematic influences cannot be overcome by longer observations and are of different origin, for example from leakage of mode power in the $l-\omega$--diagram that comes essentially from the technical restrictions in observing the solar velocity field in the photosphere and mode coupling dominated by rotation. The method used by the global helioseismic metrology project is able to reduce some of the systematic errors of the estimated parameters which provide a better agreement with solar models constructed with recent versions of the equation of state.

One systematic influence, the mode coupling, actually results from a perturbation of the mode eigenfunctions by flows in the solar interior. These perturbations lead to a cross-talk between the spherical harmonic coefficients of the Doppler velocity measurements that manifests as leakage in the power spectra. This phenomenon was recently exploited to develop global helioseismic methods to infer the meridional flow in the deeper interior. Here, the characteristic cross-talk due to the meridional flow is investigated by a cross-spectral analysis of time series of these spherical harmonic velocity coefficients. The meridional flow measured by this approach shows a complex flow pattern over latitude and depth and extends from the surface down to the base of the convection zone. Since mode eigenfunctions are also perturbed by other kinds of disturbance, e.g., rotation, their analysis may be also of interest for studies on these perturbations.  

One important objective of helioseismic investigations is to reveal the dynamic processes in the interior associated with the solar dynamo, e.g., the meridional flow, in order to better understand this mechanism. Another approach to explore this mechanism and moreover to predict the related solar magnetic activity cycle was recently made by the use of data assimilation methods. This subject was reviewed in detail in the second part of this paper. We discussed two approaches, sequential and variational data assimilation, and their application to solar data. So far, different kinds of observational quantities, for example the sunspot numbers, the speed of the meridional flow, the polar magnetic field as well as synthetic data, and rather simple models relating observations to the solar dynamo were combined to afford forecasts of the magnetic solar cycle or flares. The results of these investigations and the increasing amount of solar data of high quality make data assimilation a promising approach for forecasts of the solar activity cycle as well as magnetic activities relevant for space weather.

\begin{acknowledgements}
LJ would like to thank Sacha Brun for helpful comments on the data assimilation part in this paper. MR and AS have received funding from the European Research Council under the European Union’s Seventh Framework Program (FP/2007-2013)/ERC Grant Agreement no. 307117. The authors thank M.F. Woodard for providing Fig.~\ref{fig:schad2}.
\end{acknowledgements}

Conflict of Interest: The authors declare that they have no conflict of interest.

\end{document}